\RequirePackage{lineno}
\documentclass[aps,prc,showpacs,twocolumn,superscriptaddress]{revtex4}
\usepackage{multirow}
\usepackage{graphicx}
\usepackage{amsmath}
\usepackage{lineno}
\usepackage{color}
\usepackage{amsmath}
\usepackage{amssymb}
\usepackage{MnSymbol}
\def\version{version 6.0}
\newcommand{\sqrtsNN}{\mbox{$\sqrt{\mathrm{\it s_{NN}}}$} }
\newcommand{\vtwo}{$v_{2}$ }

\newcommand{\pt}{$p_T$ }

\def \auau  {Au + Au }
\def \cucu  {Cu + Cu }

\def \eparttwos {$\varepsilon_{\mathrm{part}}\{2\}$}
\def \GeVc {\mbox{$\mathrm{GeV}/c$}}

\begin{document}
\title{
\begin{flushright}
{
\small \sl \version \\
}
\end{flushright}
Inclusive charged hadron elliptic flow in Au + Au collisions at \sqrtsNN = 7.7 - 39 GeV
}
\affiliation{AGH University of Science and Technology, Cracow, Poland}
\affiliation{Argonne National Laboratory, Argonne, Illinois 60439, USA}
\affiliation{University of Birmingham, Birmingham, United Kingdom}
\affiliation{Brookhaven National Laboratory, Upton, New York 11973, USA}
\affiliation{University of California, Berkeley, California 94720, USA}
\affiliation{University of California, Davis, California 95616, USA}
\affiliation{University of California, Los Angeles, California 90095, USA}
\affiliation{Universidade Estadual de Campinas, Sao Paulo, Brazil}
\affiliation{Central China Normal University (HZNU), Wuhan 430079, China}
\affiliation{University of Illinois at Chicago, Chicago, Illinois 60607, USA}
\affiliation{Cracow University of Technology, Cracow, Poland}
\affiliation{Creighton University, Omaha, Nebraska 68178, USA}
\affiliation{Czech Technical University in Prague, FNSPE, Prague, 115 19, Czech Republic}
\affiliation{Nuclear Physics Institute AS CR, 250 68 \v{R}e\v{z}/Prague, Czech Republic}
\affiliation{University of Frankfurt, Frankfurt, Germany}
\affiliation{Institute of Physics, Bhubaneswar 751005, India}
\affiliation{Indian Institute of Technology, Mumbai, India}
\affiliation{Indiana University, Bloomington, Indiana 47408, USA}
\affiliation{Alikhanov Institute for Theoretical and Experimental Physics, Moscow, Russia}
\affiliation{University of Jammu, Jammu 180001, India}
\affiliation{Joint Institute for Nuclear Research, Dubna, 141 980, Russia}
\affiliation{Kent State University, Kent, Ohio 44242, USA}
\affiliation{University of Kentucky, Lexington, Kentucky, 40506-0055, USA}
\affiliation{Institute of Modern Physics, Lanzhou, China}
\affiliation{Lawrence Berkeley National Laboratory, Berkeley, California 94720, USA}
\affiliation{Massachusetts Institute of Technology, Cambridge, MA 02139-4307, USA}
\affiliation{Max-Planck-Institut f\"ur Physik, Munich, Germany}
\affiliation{Michigan State University, East Lansing, Michigan 48824, USA}
\affiliation{Moscow Engineering Physics Institute, Moscow Russia}
\affiliation{National Institute of Science and Education and Research, Bhubaneswar 751005, India}
\affiliation{Ohio State University, Columbus, Ohio 43210, USA}
\affiliation{Old Dominion University, Norfolk, VA, 23529, USA}
\affiliation{Institute of Nuclear Physics PAN, Cracow, Poland}
\affiliation{Panjab University, Chandigarh 160014, India}
\affiliation{Pennsylvania State University, University Park, Pennsylvania 16802, USA}
\affiliation{Institute of High Energy Physics, Protvino, Russia}
\affiliation{Purdue University, West Lafayette, Indiana 47907, USA}
\affiliation{Pusan National University, Pusan, Republic of Korea}
\affiliation{University of Rajasthan, Jaipur 302004, India}
\affiliation{Rice University, Houston, Texas 77251, USA}
\affiliation{Universidade de Sao Paulo, Sao Paulo, Brazil}
\affiliation{University of Science \& Technology of China, Hefei 230026, China}
\affiliation{Shandong University, Jinan, Shandong 250100, China}
\affiliation{Shanghai Institute of Applied Physics, Shanghai 201800, China}
\affiliation{SUBATECH, Nantes, France}
\affiliation{Texas A\&M University, College Station, Texas 77843, USA}
\affiliation{University of Texas, Austin, Texas 78712, USA}
\affiliation{University of Houston, Houston, TX, 77204, USA}
\affiliation{Tsinghua University, Beijing 100084, China}
\affiliation{United States Naval Academy, Annapolis, MD 21402, USA}
\affiliation{Valparaiso University, Valparaiso, Indiana 46383, USA}
\affiliation{Variable Energy Cyclotron Centre, Kolkata 700064, India}
\affiliation{Warsaw University of Technology, Warsaw, Poland}
\affiliation{University of Washington, Seattle, Washington 98195, USA}
\affiliation{Wayne State University, Detroit, Michigan 48201, USA}
\affiliation{Yale University, New Haven, Connecticut 06520, USA}
\affiliation{University of Zagreb, Zagreb, HR-10002, Croatia}

\author{L.~Adamczyk}\affiliation{AGH University of Science and Technology, Cracow, Poland}
\author{G.~Agakishiev}\affiliation{Joint Institute for Nuclear Research, Dubna, 141 980, Russia}
\author{M.~M.~Aggarwal}\affiliation{Panjab University, Chandigarh 160014, India}
\author{Z.~Ahammed}\affiliation{Variable Energy Cyclotron Centre, Kolkata 700064, India}
\author{A.~V.~Alakhverdyants}\affiliation{Joint Institute for Nuclear Research, Dubna, 141 980, Russia}
\author{I.~Alekseev}\affiliation{Alikhanov Institute for Theoretical and Experimental Physics, Moscow, Russia}
\author{J.~Alford}\affiliation{Kent State University, Kent, Ohio 44242, USA}
\author{B.~D.~Anderson}\affiliation{Kent State University, Kent, Ohio 44242, USA}
\author{C.~D.~Anson}\affiliation{Ohio State University, Columbus, Ohio 43210, USA}
\author{D.~Arkhipkin}\affiliation{Brookhaven National Laboratory, Upton, New York 11973, USA}
% ...

\collaboration{STAR Collaboration}\noaffiliation

\date{\today}

\begin{abstract}
A systematic study is presented for centrality, transverse momentum ($p_T$) and pseudorapidity ($\eta$) dependence
of the inclusive charged hadron elliptic flow ($v_2$) at midrapidity ($|\eta| < 1.0$) in Au+Au collisions at
$\sqrt{s_{NN}}$ = 7.7, 11.5, 19.6, 27 and 39 GeV. The results obtained with different methods, including correlations
with the event plane reconstructed in a region separated by a large pseudorapidity gap and 4-particle cumulants ($v_2\{4\}$),
are presented in order to investigate non-flow correlations and $v_2$ fluctuations. We observe that the difference between
$v_2\{2\}$ and $v_2\{4\}$ is smaller at the lower collision energies. Values of $v_2$,  scaled by the initial coordinate
space eccentricity, $v_{2}/\varepsilon$, as a function of $p_T$ are larger in more central collisions, suggesting stronger
collective flow develops in more central collisions, similar to the results at higher collision energies. These results are
compared to measurements at higher energies at the Relativistic Heavy Ion Collider ($\sqrt{s_{NN}}$ = 62.4 and 200 GeV) and
at the Large Hadron Collider (Pb + Pb collisions at $\sqrt{s_{NN}}$ = 2.76 TeV).
The $v_2(p_T)$ values for fixed $p_T$ rise with increasing collision energy within the $p_T$ range studied ($< 2~{\rm GeV}/c$).
A comparison to viscous hydrodynamic simulations is made to potentially help understand the energy dependence of $v_{2}(p_{T})$.
We also compare the $v_2$ results to UrQMD and AMPT transport model calculations, and physics implications on the dominance of partonic
versus hadronic phases in the system created at Beam Energy Scan (BES) energies are discussed.
\end{abstract}
\maketitle
\clearpage
\section{Introduction}
\label{sect_intro}
Azimuthal anisotropies of particle distributions relative to the reaction plane (plane subtended by the impact parameter and beam direction) in high energy heavy-ion collisions have been used to characterize the collision dynamics~\cite{flowreview1, flowreview2, flowreview3}.
In a picture of hydrodynamic expansion of the system formed in the collisions, these anisotropies are expected to arise due to initial pressure gradients and subsequent
interactions of the constituents~\cite{firsthydro, hydroBES}.
Specifically, differential measurements~\cite{star_130v2, starklv2, msv2, starwp, runII200gevV2, star_fv2, star_v2cen, cucuv2STAR,phenixv2_1,phenixprl,phobosv2,na49v2,alicev2,atlasv2}
of azimuthal anisotropy have been found to be sensitive to (a) the equation of state (EOS), (b) thermalization, (c) transport coefficients of the medium, and (d) initial conditions in the
heavy-ion collisions.
Hence it is important to study the dependence of azimuthal anisotropy as a function of several variables, for example center-of-mass energy (\sqrtsNN),
collision centrality, transverse momentum ($p_T$), and pseudorapidity ($\eta$).

Recently a beam-energy scan (BES) program has begun at RHIC to study the QCD phase diagram~\cite{bes}.
The BES program extends the baryonic chemical potential ($\mu_{B}$) reach of RHIC from 20 to about 400 MeV~\cite{stat_model, star_ub}.
The baryon chemical potential decreases with the decrease in the beam energy while the chemical freeze-out temperature increases with
increase in beam energy~\cite{chemfo}.
This allows one to study azimuthal anisotropy at midrapidity with varying net-baryon densities. Lattice QCD calculations suggest that the quark-hadron
transition is a crossover for high temperature ($T$) systems with small $\mu_{B}$ or high \sqrtsNN~\cite{lattice}.
Several model calculations suggest that at larger values of $\mu_{B}$ or lower \sqrtsNN the transition is expected to be first order~\cite{order1,order2,order}.
Theoretical calculations suggest a non-monotonic behavior of $v_2$ could be observed around this ``softest point of the EOS''~\cite{CP_predict}.
The softest point of the EOS is usually referred to as the temparature/time during which the velocity of sound has a minimum value (or reduction in
the pressure of the system) during the evolution. Non-monotonic variation of azimuthal anisotropy as a function of collision
centrality and \sqrtsNN  could indicate the softest point of the EOS in heavy-ion reactions~\cite{sorge}.
Further it has been argued that the observation of saturation of differential azimuthal anisotropies $v_2(p_T)$ of charged hadrons
in \auau collisions in the \sqrtsNN range of 62.4 - 200 GeV is a signature of a mixed phase~\cite{phenixprl}.
The new data presented in this paper shows to what extent such a saturation effect is observed.

Several analysis methods for $v_2$ have been proposed~\cite{v2Methods,2part,cumulant1, cumulant2, LYZ}. These are found to
be sensitive in varying degrees to non-flow contributions (e.g. correlations due to jets, resonances, etc.) and flow fluctuations.
$v_{2}$ measurements from various methods have been judiciously used to constrain these contributions, in addition to providing
estimates of systematic errors associated with the
measurements~\cite{flowfluc}. This is  particularly useful for interpreting results of identified hadron
$v_{2}$ values where, due to limitations of event statistics, it is not possible to use all methods for $v_{2}$ analysis.
The measurements over a range of energies may provide insights to the evolution of non-flow and
flow fluctuations as a function of collision energy.

Inclusive charged hadron elliptic flow measurements at top RHIC energies have been one of
the most widely studied observables from the theoretical perspective. It has been shown that transport models,
which provide a microscopic description of the early and late non-equilibrium stages of the system,
significantly underpredict $v_2$ at top RHIC energies, while the inclusion of partonic effects provides
a more satisfactory explanation~\cite{nasim}.
The new data discussed here will provide an opportunity to study the contribution of partonic matter
and hadronic matter to the $v_2$ measurements as a function  of \sqrtsNN or ($T$, $\mu_{\mathrm B}$)
by comparisons with models.

In this paper we present measurements of the second harmonic azimuthal anisotropy  using data taken in the BES program
from \sqrtsNN = 7.7 to 39 GeV. We discuss the detectors used in the analysis, data selections and
methods used to determine inclusive charged hadron $v_{2}$ in Sections II and III. Section IV gives $v_2$ results for inclusive charged hadrons from different
analysis methods. We discuss the centrality, $\eta$, $p_{T}$ and \sqrtsNN dependence of $v_{2}$ in Section V, and compare to calculations from transport models.
Finally, a summary of the analysis is presented in Section VI.

\section{Experiments and data sets}
\subsection{STAR detector}
The results presented here are based on data collected during the tenth and eleventh RHIC runs (2010 and 2011) with the STAR detector using minimum-bias triggers
(requiring a combination of signals from the Beam-Beam Counters (BBC)~\cite{BBC}, Zero Degree Calorimeters (ZDC)~\cite{ZDC}, and Vertex Position Detectors (VPD)~\cite{VPD}).
For the 7.7 and 11.5 GeV data, at least one hit in the full barrel Time-Of-Flight detector~\cite{tof} was required in order to further reduce the background. The main Time Projection Chamber
(TPC)~\cite{startpc} and two Forward Time Projection Chambers (FTPCs)~\cite{starftpc} were used for particle tracking in the central region ($|\eta| < 1.0$) and forward regions ($2.5 < |\eta| < 4.0$) respectively.
Both the TPC and FTPCs provided azimuthal acceptance over $2\pi$.
The BBC detector subsystem consists of two detectors mounted around the beam pipe, each located outside the STAR magnet pole-tip at
opposite ends of the TPC approximately 375 cm from the center of the nominal interaction point.  Each BBC detector consists of hexagonal scintillator tiles arranged in four concentric rings that provided full azimuthal coverage. The inner tiles of the BBCs, with a pseudorapidity range of $3.8 < |\eta| < 5.2$ were used to reconstruct the event plane in one elliptic flow analysis.

\subsection{Event and track selection}
Events for analysis are selected based on collision vertex positions within 2 cm of the beam axis to reduce contributions from beam-gas and beam-pipe (at a radius of 4 cm) interactions,
and within a limited distance from the center of the detector along the beam direction ($\pm70$ cm for the 7.7 GeV data set, $\pm50$ cm for the 11.5 GeV data set, and $\pm40$ cm for the 19.6, 27 and 39 GeV data sets).
These values are chosen to reduce systematics due to variance in detector performance over $|\eta| < 1.0$ while retaining sufficient statistics.
After quality cuts, about 4 million $0-80\%$ central events remain for 7.7 GeV, 11 million for 11.5 GeV, 20 million for 19.6 GeV, 40 million for 27 GeV and 120 million for 39 GeV data sets.
The results from more peripheral collisions are not presented due to trigger inefficiencies at low multiplicity.
The centrality was defined using the number of charged tracks with quality cuts similar to those in Ref.~\cite{star_v2cen}.
The details of the centrality determination will be discussed in subsection C.
The $0-80\%$ central events for \vtwo analysis of charged hadrons are divided into nine centrality bins: $0-5\%$, $5-10\%$, $10-20\%$, $20-30\%$, $30-40\%$, $40-50\%$, $50-60\%$, $60-70\%$ and $70-80\%$.

A variety of track quality cuts are used to select good charged particle tracks reconstructed using information from the TPC or FTPCs.
The distance of closest approach (DCA) of the track to the primary vertex is taken to be less than 2 cm.
We require that the TPC and FTPCs have a number of fit points used for reconstruction of the tracks
to be $> 15$ and $> 5$, respectively. For the TPC and FTPCs the ratio of the number of fit points to maximum possible hits
is $> 0.52$. An additional transverse momentum cut ($0.2 < p_{T} < 2\ \GeVc$) is applied to the charged tracks for
the TPC and FTPC event plane determination.


\subsection{Centrality determination}

\begin{figure*}[ht]
\vskip 0cm
\includegraphics[width=0.7\textwidth]{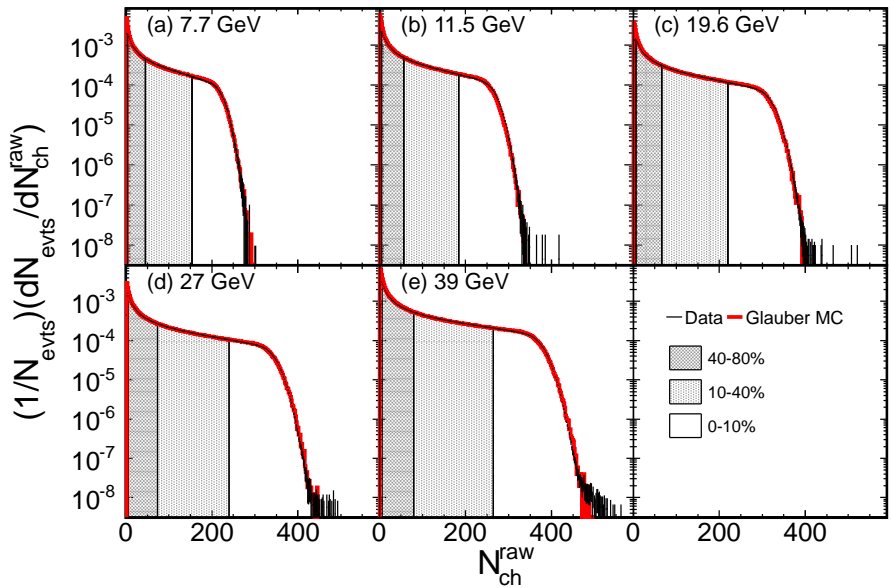}
\caption{ (Color online) Distribution of uncorrected multiplicity $N_{\rm ch}^{\rm raw}$ measured within $|\eta| <$~0.5 in the TPC from  \sqrtsNN = 7.7 to 39 GeV
in Au + Au collisions shown as black points. The red curves show the multiplicity distributions at \sqrtsNN = 7.7 to 39 GeV from MC Glauber simulations.
See texts for more details about simulations.
} \label{fig:centrality}
\end{figure*}

The centrality classes are defined based on the uncorrected charged particle multiplicity ($N_{\rm ch}^{\rm raw}$) distribution
in the TPC for pseudorapidity $|\eta| <$ 0.5 and full azimuth.

Figure~\ref{fig:centrality} shows the $N_{\rm ch}^{\rm raw}$ distribution for charged particles from the data at \sqrtsNN = 7.7, 11.5, 19.6, 27 and 39 GeV
compared to those from Monte Carlo (MC) Glauber simulations.
The detailed procedures to obtain the simulated multiplicity are similar to that described
in~\cite{centralitydef}. A two-component model~\cite{Kharzeev:2000ph} is used to calculate the simulated
multiplicity distribution given by
\begin{equation}
\frac{dN_{\rm ch}}{d\eta}\bigg|_{\eta = 0}= n_{pp}\left[(1-x)\frac{N_{\rm part}}{2} + xN_{\rm coll}\right],
\end{equation}
where $N_{\rm part}$ is the number of participant nucleons and $N_{\rm coll}$ is the number of binary nucleon-nucleon collisions
in the simulations. The fitting parameter $n_{pp}$ is the average multiplicity per unit of pseudorapidity in
minimum-bias $p$ + $p$ collisions and $x$ is the fraction of production from the hard component. The inelastic nucleon-nucleon
cross section $\sigma_{NN}^{\rm inel}$ is extracted from fitting the results of available data for total and
elastic $p$ + $p$ cross sections from the Particle Data Group~\cite{Nakamura:2010zzi}.
The $x$ value is fixed at 0.12 $\pm$ 0.02 based on the
linear interpolation of the PHOBOS results at \sqrtsNN = 19.6 and 200 GeV~\cite{Back:2004dy}.
Systematic errors on $n_{pp}$ are evaluated by varying both $n_{pp}$ and $x$ within the quoted $x$ uncertainty
to determine the minimum $\chi^2$ to describe the data. Since the $n_{pp}$ and $x$ are anti-correlated,
lower (higher) $n_{pp}$ is used for higher (lower) $x$ for systematic error evaluations on $N_{\rm part}$.
Table~\ref{tab:twocomponent_parameters} summarizes the parameters in the two-component model and
$\sigma_{NN}^{\rm inel}$ in the MC Glauber simulations. The event-by-event multiplicity fluctuations are
included using negative binomial distributions~\cite{centralitydef}. The centrality classes are defined
by the fractions of geometrical cross section from the simulated multiplicity distributions.
For each centrality bin, average quantities are calculated in the MC Glauber simulations for $\left<N_{\rm part}\right>$, $\left<N_{\rm coll}\right>$,
reaction plane eccentricity $\left<\varepsilon_{\rm RP}\right>$, participant eccentricity $\left<\varepsilon_{\rm part}\right>$,
root-mean-square participant eccentricity $\varepsilon_{\rm part}\{2\}$, and transverse area
$\left<S_{\rm part}\right>$. Eccentricity and transverse area are defined by
\begin{eqnarray}
\varepsilon_{\rm RP} & = & \frac{\sigma_{y}^2 - \sigma_{x}^2}{\sigma_{x}^2 + \sigma_{y}^2}, \\
\varepsilon_{\rm part}    & = &  \frac{\sqrt{(\sigma_y^2-\sigma_x^2)^2 + 4\sigma_{xy}^2}}{\sigma_x^2+\sigma_y^2},
\varepsilon_{\rm part}\{2\} = \sqrt{\langle\varepsilon_{\rm part}^{2}\rangle}, \\
S_{\rm part}             & = & \pi\sqrt{\sigma_x^2\sigma_y^2 - \sigma_{xy}^2}, \\
\sigma_x^2           & = & \{x^2\} - \{x\}^2, ~ \sigma_y^2 = \{y^2\} - \{y\}^2, \\
\sigma_{xy}          & = & \{xy\} - \{x\}\{y\},
\end{eqnarray}
where the curly brackets denote the average over all participants per event, and $x$ and $y$ are the positions of
participant nucleons. Systematic uncertainties on those quantities are evaluated by varying parameters for the
two-component model and by varying the input parameters in the MC Glauber model. The quoted errors are the
quadratic sum of the individual systematic uncertainties.
\begin{table}[ht]
\caption{\label{tab:twocomponent_parameters}
Summary of $n_{pp}$ and $\sigma_{NN}^{\rm inel}$ with
systematic uncertainties at \sqrtsNN = 7.7, 11.5, 19.6, 27 and 39 GeV.
$x$ is set to 0.12 $\pm$ 0.02 for all collision energies. }
\begin{tabular}{cp{6em}p{7em}p{6em}}
\hline
\sqrtsNN (GeV) & $n_{pp}$ & $\sigma_{NN}^{\rm inel}$ (mb)\\
\hline
7.7    &  0.89 $\pm$ 0.04 &  30.8 $\pm$ 1.0 \\
11.5   &  1.07 $\pm$ 0.05 &  31.2 $\pm$ 1.0 \\
19.6   &  1.29 $\pm$ 0.05 &  32.0 $\pm$ 1.0 \\
27     &  1.39 $\pm$ 0.06 &  33.0 $\pm$ 1.0 \\
39     &  1.52 $\pm$ 0.08 &  34.0 $\pm$ 1.0 \\
\hline
\end{tabular}
\end{table}
Table~\ref{tab:glauber} summarizes the centrality classes as well as the results obtained by MC Glauber
simulations at the five energies.

\begin{table*}[ht]
\caption{
Summary of centrality bins, average number of participants $\left<N_{\rm part}\right>$,
number of binary collisions $\left<N_{\rm coll}\right>$,
reaction plane eccentricity $\left<\varepsilon_{\rm RP}\right>$,
participant eccentricity $\left<\varepsilon_{\rm part}\right>$,
root-mean-square the participant eccentricity $\varepsilon_{\rm part}\{2\}$
and transverse area $\left<S_{\rm part}\right>$
from MC Glauber simulations at \sqrtsNN = 7.7, 11.5, 19.6, 27, and 39 GeV.
The errors are systematic uncertainties.
}
\begin{tabular}{cp{8em}p{8em}p{8em}p{8em}p{8em}p{8em}}
\hline \hline
Centrality (\%) & $\left<N_{\rm part}\right>$ & $\left<N_{\rm coll}\right>$ & $\left<\varepsilon_{\rm RP}\right>$ & $\left<\varepsilon_{\rm part}\right>$ & $\varepsilon_{\rm part}\{2\}$ & $\left<S_{\rm part}\right>$ (fm$^2$) \\
\hline
\multicolumn{7}{c}{\auau at \sqrtsNN = 7.7 GeV} \\
\hline
0-5\% & 337 $\pm$ 2  & 774 $\pm$ 28 & 0.043 $\pm$ 0.007 & 0.102 $\pm$ 0.003 & 0.117 $\pm$ 0.003 & 25.5 $\pm$ 0.4 \\
5-10\% & 290 $\pm$ 6  & 629 $\pm$ 20 & 0.10  $\pm$  0.01 & 0.14 $\pm$ 0.01  & 0.16 $\pm$ 0.01   & 23.0 $\pm$ 0.3 \\
10-20\% & 226 $\pm$ 8  & 450 $\pm$ 22 & 0.18  $\pm$  0.02 & 0.21 $\pm$ 0.02 & 0.24 $\pm$ 0.02   & 19.5 $\pm$ 0.4 \\
20-30\% & 160 $\pm$ 10 & 283 $\pm$ 24 & 0.26  $\pm$  0.03 & 0.30 $\pm$ 0.02 & 0.32 $\pm$ 0.02    & 15.7 $\pm$ 0.7 \\
30-40\% & 110 $\pm$ 11 & 171 $\pm$ 23 & 0.32  $\pm$  0.04 & 0.37 $\pm$ 0.03 & 0.39 $\pm$ 0.03   & 12.6 $\pm$ 0.8 \\
40-50\% &  72 $\pm$ 10 &  96 $\pm$ 19 & 0.36  $\pm$  0.04 & 0.43 $\pm$ 0.03 & 0.46 $\pm$ 0.03   &  10.0 $\pm$ 0.9 \\
50-60\% &  45 $\pm$ 9  &  52 $\pm$ 13 & 0.39  $\pm$  0.04 & 0.50 $\pm$ 0.03 & 0.53 $\pm$ 0.03   &  7.8 $\pm$ 1.0 \\
60-70\% &  26 $\pm$ 7  &  25 $\pm$  9 & 0.40  $\pm$  0.05 & 0.58 $\pm$ 0.04 & 0.62 $\pm$ 0.04   &  5.8 $\pm$ 1.1 \\
70-80\% &  14 $\pm$ 4  &  12 $\pm$  5 & 0.36  $\pm$  0.05 & 0.68 $\pm$ 0.04 & 0.72 $\pm$ 0.04   &  3.6 $\pm$ 1.0 \\
\hline
\multicolumn{7}{c}{\auau at \sqrtsNN = 11.5 GeV} \\
\hline
0-5\% & 338 $\pm$ 2  & 784 $\pm$ 27 & 0.043 $\pm$ 0.006 & 0.102 $\pm$ 0.003 & 0.116 $\pm$ 0.003 & 25.6 $\pm$ 0.4 \\
5-10\% & 290 $\pm$ 6  & 635 $\pm$ 20 & 0.10  $\pm$  0.01 & 0.14 $\pm$ 0.01  & 0.16 $\pm$ 0.01  & 23.0 $\pm$ 0.3 \\
10-20\% & 226 $\pm$ 8  & 453 $\pm$ 23 & 0.18  $\pm$  0.02 & 0.22 $\pm$ 0.02 & 0.24 $\pm$ 0.02   & 19.5 $\pm$ 0.5 \\
20-30\% & 160 $\pm$ 9 & 284 $\pm$ 23 & 0.26  $\pm$  0.03 & 0.30 $\pm$ 0.02 & 0.32 $\pm$ 0.02   & 15.7 $\pm$ 0.7 \\
30-40\% & 110 $\pm$ 10 & 172 $\pm$ 22 & 0.32  $\pm$  0.04 & 0.37 $\pm$ 0.03 & 0.39 $\pm$ 0.03   & 12.6 $\pm$ 0.8 \\
40-50\% &  73 $\pm$ 10 & 98 $\pm$ 18 & 0.36  $\pm$  0.04 & 0.43 $\pm$ 0.03 & 0.46 $\pm$ 0.03   &  10.1 $\pm$ 0.9 \\
50-60\% &  44 $\pm$ 9  &  52 $\pm$ 14 & 0.39  $\pm$  0.04 & 0.50 $\pm$ 0.03 & 0.53 $\pm$ 0.03   &  7.8 $\pm$ 1.0 \\
60-70\% &  26 $\pm$ 7  &  25 $\pm$  9 & 0.40  $\pm$  0.05 & 0.58 $\pm$ 0.04 & 0.62 $\pm$ 0.04   &  5.8 $\pm$ 1.1 \\
70-80\% &  14 $\pm$ 6  &  12 $\pm$  6 & 0.37  $\pm$  0.06 & 0.68 $\pm$ 0.05 & 0.71 $\pm$ 0.05   &  3.7 $\pm$ 1.2 \\\hline
\multicolumn{7}{c}{\auau at \sqrtsNN = 19.6 GeV} \\
\hline
0-5\% & 338 $\pm$ 2  & 800 $\pm$ 27 & 0.044 $\pm$ 0.006 & 0.102 $\pm$ 0.003 & 0.117 $\pm$ 0.003 & 25.6 $\pm$ 0.4 \\
5-10\% & 289 $\pm$ 6  & 643 $\pm$ 20 & 0.11  $\pm$  0.01 & 0.15 $\pm$ 0.01  & 0.16 $\pm$ 0.01  & 23.0 $\pm$ 0.3 \\
10-20\% & 225 $\pm$ 9  & 458 $\pm$ 24 & 0.18  $\pm$  0.02 & 0.22 $\pm$ 0.02 & 0.24 $\pm$ 0.02   & 19.5 $\pm$ 0.5 \\
20-30\% & 158 $\pm$ 10 & 284 $\pm$ 26 & 0.26  $\pm$  0.03 & 0.30 $\pm$ 0.02 & 0.32 $\pm$ 0.02   & 15.6 $\pm$ 0.7 \\
30-40\% & 108 $\pm$ 10 & 170 $\pm$ 23 & 0.32  $\pm$  0.04 & 0.37 $\pm$ 0.03 & 0.40 $\pm$ 0.03   & 12.5 $\pm$ 0.8 \\
40-50\% &  71 $\pm$ 10 & 96 $\pm$ 18 & 0.36  $\pm$  0.04 & 0.43 $\pm$ 0.03 & 0.46 $\pm$ 0.03   & 10.0 $\pm$ 0.9 \\
50-60\% &  44 $\pm$ 9  &  51 $\pm$ 13 & 0.39  $\pm$  0.04 & 0.50 $\pm$ 0.03 & 0.53 $\pm$ 0.03   &  7.8 $\pm$ 1.0 \\
60-70\% &  25 $\pm$ 7  &  25 $\pm$  8 & 0.40  $\pm$  0.05 & 0.58 $\pm$ 0.04 & 0.62 $\pm$ 0.04   & 5.8 $\pm$ 1.1 \\
70-80\% &  14 $\pm$ 5  &  12 $\pm$  5 & 0.37  $\pm$  0.06 & 0.68 $\pm$ 0.05 & 0.71 $\pm$ 0.05   &  3.7 $\pm$ 1.2 \\\hline
\multicolumn{7}{c}{\auau at \sqrtsNN = 27 GeV} \\
\hline
0-5\% & 343 $\pm$ 2  & 841 $\pm$ 28 & 0.040 $\pm$ 0.005 & 0.100 $\pm$ 0.002 & 0.114 $\pm$ 0.003 & 25.8 $\pm$ 0.4 \\
5-10\% & 299 $\pm$ 6  & 694 $\pm$ 22 & 0.10  $\pm$  0.01 & 0.14 $\pm$ 0.01  & 0.16 $\pm$ 0.01  & 23.4 $\pm$ 0.3 \\
10-20\% & 233 $\pm$ 9  & 497 $\pm$ 26 & 0.18  $\pm$  0.02 & 0.21 $\pm$ 0.02 & 0.23 $\pm$ 0.02   & 19.8 $\pm$ 0.5 \\
20-30\% & 166 $\pm$ 11 & 312 $\pm$ 28 & 0.26  $\pm$  0.03 & 0.29 $\pm$ 0.02 & 0.32 $\pm$ 0.02   & 15.9 $\pm$ 0.7 \\
30-40\% & 114 $\pm$ 11 & 188 $\pm$ 25 & 0.32  $\pm$  0.04 & 0.37 $\pm$ 0.03 & 0.39 $\pm$ 0.03   & 12.8 $\pm$ 0.9 \\
40-50\% &  75 $\pm$ 10 & 106 $\pm$ 20 & 0.37  $\pm$  0.04 & 0.43 $\pm$ 0.03 & 0.46 $\pm$ 0.03   & 10.2 $\pm$ 0.9 \\
50-60\% &  47 $\pm$ 9  &  56 $\pm$ 15 & 0.39  $\pm$  0.05 & 0.50 $\pm$ 0.03 & 0.53 $\pm$ 0.03   &  7.9 $\pm$ 1.0 \\
60-70\% &  27 $\pm$ 8  &  27 $\pm$  10 & 0.40  $\pm$  0.05 & 0.58 $\pm$ 0.05 & 0.61 $\pm$ 0.05   &  5.8 $\pm$ 1.2 \\
70-80\% &  14 $\pm$ 6  &  12 $\pm$  6 & 0.37  $\pm$  0.06 & 0.68 $\pm$ 0.05 & 0.71 $\pm$ 0.05   &  3.6 $\pm$ 1.3 \\
\hline
\multicolumn{7}{c}{\auau at \sqrtsNN = 39 GeV} \\
\hline
0-5\% & 342 $\pm$ 2  & 853 $\pm$ 27 & 0.042 $\pm$ 0.006 & 0.101 $\pm$ 0.003 & 0.115 $\pm$ 0.003 & 25.9 $\pm$ 0.4 \\
5-10\% & 294 $\pm$ 6  & 687 $\pm$ 21 & 0.10  $\pm$  0.01 & 0.14 $\pm$ 0.01  & 0.16 $\pm$ 0.01  & 23.3 $\pm$ 0.3 \\
10-20\% & 230 $\pm$ 9  & 492 $\pm$ 26 & 0.18  $\pm$  0.02 & 0.21 $\pm$ 0.02 & 0.23 $\pm$ 0.02   & 19.8 $\pm$ 0.5 \\
20-30\% & 162 $\pm$ 10 & 306 $\pm$ 27 & 0.26  $\pm$  0.03 & 0.30 $\pm$ 0.02 & 0.32 $\pm$ 0.02   & 16.0 $\pm$ 0.7 \\
30-40\% & 111 $\pm$ 11 & 183 $\pm$ 24 & 0.32  $\pm$  0.04 & 0.37 $\pm$ 0.03 & 0.39 $\pm$ 0.03   & 12.8 $\pm$ 0.8 \\
40-50\% &  74 $\pm$ 10 & 104 $\pm$ 20 & 0.36  $\pm$  0.04 & 0.43 $\pm$ 0.03 & 0.46 $\pm$ 0.03   & 10.3 $\pm$ 1.0 \\
50-60\% &  46 $\pm$ 9  &  55 $\pm$ 14 & 0.39  $\pm$  0.04 & 0.50 $\pm$ 0.03 & 0.53 $\pm$ 0.03   &  8.0 $\pm$ 1.0 \\
60-70\% &  26 $\pm$ 7  &  27 $\pm$  9 & 0.40  $\pm$  0.04 & 0.58 $\pm$ 0.04 & 0.61 $\pm$ 0.04   &  5.9 $\pm$ 1.1 \\
70-80\% &  14 $\pm$ 5  &  12 $\pm$  6 & 0.37  $\pm$  0.05 & 0.67 $\pm$ 0.05 & 0.71 $\pm$ 0.05   &  3.8 $\pm$ 1.2 \\
\hline \hline

\end{tabular}
\label{tab:glauber}
\end{table*}

\section{Elliptic flow methods}
\label{flow_method}
\subsection{The event plane method}
The event plane method~\cite{v2Methods} correlates each particle with the event plane determined from the full event
minus the particle of interest, which can be done for each harmonic. For any Fourier harmonic, $n$,  the event flow vector
($Q_n$) and the event plane angle ($\Psi_n$) are defined by~\cite{v2Methods}
\begin{equation}
Q_n \cos n\Psi_n  = Q_{nx} = \sum_i w_i\cos n\phi_i,
\label{Qx}
\end{equation}
\begin{equation}
Q_n \sin n\Psi_n = Q_{ny} = \sum_i w_i\sin n\phi_i,
\label{Qy}
\end{equation}
\begin{equation}
\Psi_n  =  \left( \tan^{-1} \frac{Q_{ny}}{Q_{nx}} \right)/n,
\end{equation}
where sums extend over all particles $i$ used in the event plane calculation, and $\phi_i$ and $w_i$ are the laboratory
azimuthal angle and the weight for the $i^{\rm th}$ particle, respectively. The reaction plane azimuthal distribution
should be isotropic or flat in the laboratory frame if the detectors
have ideal acceptance. Since the detectors usually have non-uniform acceptance, a procedure for flattening the laboratory event
plane distribution is necessary ~\cite{recenter, shiftMethod}.

As shown in Eq.~(\ref{v2obs}), the observed $v_2$ is calculated with respect to the reconstructed event plane angle
$\Psi_{n}$ where $n$ equals 2 when we use the second harmonic event plane and $n$ equals  1 when we use the first harmonic event plane.
\begin{equation} \label{v2obs} v_{2}^{\mathrm{obs}}\ =\ \langle
\cos[2(\phi-\Psi_n)]\rangle
\end{equation}
The angular brackets indicate an average over all particles in all events. However, tracks used for the $v_2$ calculation are excluded
from the calculation of the flow vector to remove self-correlation effects.
Because the estimated reaction plane fluctuates due to finite number of particles, one has to correct for this smearing by dividing the observed correlation by the event-plane
resolution (the denominator in Eq.~(\ref{v2EP2})), which is the correlation of the event plane with the reaction plane.
\begin{equation} \label{v2EP2} v_{2}\ =\
\frac{v_2^{\mathrm{obs}}}{\langle \cos[2(\Psi_n-\Psi_r)]\rangle}
\end{equation}

Since the reaction plane is unknown, the denominator in Eq.~(\ref{v2EP2}) could not be calculated directly. As shown in
Eq.~(\ref{EPres}), we estimate the event plane resolution by the correlation between the azimuthal angles of two subset groups of
tracks, called sub-events $A$ and $B$. In Eq.~(\ref{EPres}) $C$ is a factor calculated from the known multiplicity dependence
of the resolution~\cite{v2Methods}.
\begin{equation}
\langle \cos[2(\Psi_n-\Psi_r)]\rangle\ =\ C \sqrt { \langle
\cos[2(\Psi_{n}^{A}-\Psi_{n}^{B})] \rangle} \label{EPres}
\end{equation}
Random sub-events are used for TPC event plane, while pseudorapidity sub-events are
used for FTPC/BBC event plane.

\subsubsection{TPC event plane}
The TPC event plane means the event plane reconstructed from tracks recorded by the TPC.
For this event plane the $\phi$ weight method is an effective way to
flatten the azimuthal distribution for removing detector acceptance bias. These weights are generated by inverting the $\phi$ distributions of
detected tracks for a large event sample. The $\phi$ weights are folded into the weight $w_i$ in Eq.~(\ref{Qx}) and Eq.~(\ref{Qy}).

The re-centering correction~\cite{recenter, shiftMethod} is another method to calibrate the event plane.
In this method, one subtracts from the Q-vector of each event
the Q-vector averaged over many events. For both the $\phi$ weight and re-centering methods, the corrections
are applied in each centrality bin, in 2 bins of the primary vertex position along the longitudinal beam direction
($V_z$), and in 2 bins for positive/negative pseudorapidity. These corrections are determined as a function of data collection time.
The difference in the effects on $v_2$ from the different flattening techniques is negligible.

\subsubsection{FTPC event plane}
\label{sect_FTPCEP}
Forward-going tracks reconstructed in the two FTPCs can also be used to determine the event plane.
However, large acceptance losses from hardware faults caused
significant gaps in the azimuthal angle distribution of these tracks, preventing
use of the $\phi$ weight method because of the inability to define $\phi$ weights in regions of zero acceptance.
Thus, only the re-centering method is used for the FTPC.

\subsubsection{BBC event plane}
In this method the first-order event plane is reconstructed
using particle trajectories determined from hits in the BBC detectors. In this case,  $\phi_i$ denotes the fixed azimuthal
angle of the center of the $i^{\rm th}$ BBC tile in Eq.~(\ref{Qx}) and (\ref{Qy}), and $w_i$ is the fraction of
BBC-observed energy deposition recorded in tile $i$:
\begin{equation}
 w_i = \frac{A_{i}}{\sum A_{i}}.
\end{equation}

The BBC event plane obtained from one BBC detector is called a sub-event.  A combination of the sub-event plane vectors for
both BBC detectors provides the full event plane.
\begin{equation}
v_{2}\{{\rm BBC}\}\   = \frac{\langle \cos[2(\phi - \Psi_{1})]\rangle}{C \sqrt{\langle \cos[2(\Psi_{1}^{A} - \Psi_{1}^{B})]\rangle}}
\label{bbc}
\end{equation}
where $C$ is the constant in Eq.~(\ref{EPres}).  $ \Psi_{1}^{A}$ , $ \Psi_{1}^{B}$ are sub-event plane angles from
each BBC detector and $\Psi_{1}$ is the full event plane angle from both sub-events combined.

The detector acceptance bias is removed by applying the shift method~\cite{shiftMethod}. Equation~(\ref{shift}) shows the formula for
the shift correction. The averages in Eq.~(\ref{shift}) are taken from a large sample of events. In this analysis, the correction is done up
to the twentieth harmonic. The distributions of $\Psi_{1}^{\rm A}$ and $\Psi_{1}^{\rm B}$ are separately flattened and then
the full-event event plane distribution is flattened. Accordingly, the observed \vtwo and resolution are calculated using the shifted
(sub)event plane azimuthal angles.
\begin{equation}
\displaystyle
\begin{array}{ll}
\Psi^{'}\  & = \ \Psi \ +\ \sum_{n}\frac{1}{n}[-\langle \sin(2n\Psi)
\rangle \cos(2n\Psi) \\
\\
 & +\ \langle \cos(2n\Psi) \rangle \sin(2n\Psi)]
\end{array}
\label{shift}
\end{equation}
More details for the BBC event plane have been described in Ref.~\cite{BBCEP}.

\subsection{The $\eta$ sub-event method}
The $\eta$ sub-event method is similar to the event plane method, except one defines the flow vector for each particle based
on particles measured in the opposite hemisphere in pseudorapidity:
\begin{equation}
v_{2}\{\rm {EtaSubs}\}\   = \frac{\langle \cos[2(\phi_{\pm} - \Psi_{2, \eta_{\mp}})]\rangle}{\sqrt{\langle \cos[2(\Psi_{2, \eta_{+}} - \Psi_{2, \eta_{-}})]\rangle}}
\label{etasub}
\end{equation}

Here $v_{2}\{\rm {EtaSubs}\}$ denotes the results of the $\eta$ sub-event method and
$\Psi_{2, \eta_{+}}$($\Psi_{2, \eta_{-}}$)
is the second harmonic event plane angle determined by particles with positive (negative) pseudorapidity.
An $\eta$ gap of $|\eta| < 0.075$ is used between negative (positive) $\eta$ sub-event to reduce
non-flow correlations between the two ensembles.

\subsection{The cumulant method}
\label{sect_cumulantmethod}
The advantage of the cumulant method is that the multi-particle cumulant is a higher-order multi-particle correlation formalism
which removes the contribution of non-flow correlations from lower-order correlations~\cite{cumulant1, cumulant2}.
The measured 2-particle correlations can be expressed with flow and non-flow components:
\begin{equation}
\begin{array}{ll}
\ \langle e^{in(\phi_1 - \phi_2)} \rangle\ & = \ \langle e^{in(\phi_1 - \Psi_r)} \rangle \langle e^{in(\Psi_r - \phi_2)} \rangle + \delta_n \\
& = \ v_n^2 + \delta_n\
\end{array}
\label{twoparticle}
\end{equation}
Here $n$ is the harmonic number and $\delta_n$ denotes the non-flow contribution. The average should be taken for all pairs
of particles in a certain rapidity and transverse momentum region, and for all events of a data sample. The measured
4-particle correlations can be expressed as:
\begin{equation}
\begin{array}{ll}
\ \langle e^{in(\phi_1 + \phi_2 - \phi_3 - \phi_4)} \rangle\ & = \ v_n^4 + 2 \cdot 2 \cdot v_n^2 \delta_n + 2 \delta_n^2 \\
 \end{array}
\label{fourparticle}
\end{equation}
Thus the flow contribution can be obtained by subtracting the 2-particle correlation from the 4-particle correlation:
\begin{equation}
\begin{array}{ll}
\ \langle \langle e^{in(\phi_1 + \phi_2 - \phi_3 - \phi_4)} \rangle\rangle\ & = \ \langle e^{in(\phi_1 + \phi_2 - \phi_3 - \phi_4)} \rangle - \\
& 2\langle e^{in(\phi_1  - \phi_3 )} \rangle^{2} = \ -v_n^4\
\end{array}
\label{fourparticleCum}
\end{equation}
where $\langle \langle ... \rangle\rangle$ is used for the cumulant. The cumulant of order two is just
$\langle \langle e^{in(\phi_1 - \phi_2)} \rangle\rangle = \langle e^{in(\phi_1 - \phi_2)} \rangle$.

\subsubsection{The cumulant method with generating function}
The GF-cumulant method is computed from a generating function~\cite{cumulant2}:
\begin{equation}
G_n(z)\   = \prod_{j=1}^{M}[1 + \frac{w_j}{M}(z^{\ast}e^{in\phi_j}+ze^{-in\phi_j})]
\label{GF}
\end{equation}
Here $z$ is an arbitrary complex number, $z^{\ast}$ denotes its complex conjugate, $M$ denotes the
multiplicity in each event, and $w_j$ is the weight (transverse momentum, rapidity etc.) used in the analysis.
The event-wise averaged generating function then can be expanded in powers of  $z$ and $z^{\ast}$ where the coefficients of
expansion yield the correlations of interest:
\begin{equation}
\begin{array}{ll}
\displaystyle
\ \langle G_{n}(z)\rangle\ & = \ 1 + z\langle e^{-in\phi_1}\rangle + z^{\ast}\langle e^{in\phi_1}\rangle + \\
& \frac{M-1}{M}(\frac{z^2}{2}\langle e^{-in(\phi_1 + \phi_2)}\rangle + \frac{z^{\ast2}}{2}\langle e^{in(\phi_1 + \phi_2)}\rangle\\
& + zz^{\ast}\langle e^{in(\phi_1 - \phi_2)} \rangle) + ...
 \end{array}
\label{averageGF}
\end{equation}
These correlations can be used to construct the cumulants. More details for the analysis
of STAR data have been described in Ref.~\cite{runII200gevV2}.

\subsubsection{The Q-cumulants method}

The Q-cumulants method~\cite{Bilandzic:2010jr} is a recent method to calculate cumulants without using nested loops over
tracks and without generating functions~\cite{cumulant2}. The advantage is that it provides fast (one loop over data)
and exact non-biased (no approximations and no interference between different harmonics) estimates of the correlators
compared to the generating function cumulants. The cumulants are expressed in terms of the moments of the magnitude of the
corresponding flow vector $Q_n$
\begin{equation}
Q_n \equiv \sum_{i=1}^{M} e^{i n \phi_i}
\end{equation}
The single-event average two- and four-particle azimuthal correlations can be then formulated as:
\begin{equation}
\langle 2 \rangle = \frac{|Q_n|^2 - M}{M (M-1)}
\end{equation}
\begin{equation}
\begin{split}
\langle 4 \rangle = & \frac{|Q_n|^4 + |Q_{2n}|^2 -
2 \cdot \Re \left[Q_{2n} Q_n^* Q_n^*\right]}{M(M-1)(M-2)(M-3)} \\
& - 2 \frac{2(M-2) \cdot |Q_n|^2 - M(M-3)}{M(M-1)(M-2)(M-3)}
\end{split}
\end{equation}
The average over all events can be performed as:
\begin{equation}
\begin{split}
\llangle 2 \rrangle & \equiv \llangle e^{i n (\phi_1 - \phi_2)} \rrangle \\
                    & \equiv \frac{\sum_{{\rm event}}\left(W_{\langle 2 \rangle}\right)_i \langle 2 \rangle_i}{\sum_{{\rm event}}\left( W_{\langle 2 \rangle} \right)_i}
\end{split}
\end{equation}
\begin{equation}
\begin{split}
\llangle 4 \rrangle & \equiv \llangle e^{i n (\phi_1 + \phi_2 - \phi_3 - \phi_4)} \rrangle \\
                    & \equiv \frac{\sum_{{\rm events}}\left(W_{\langle 4 \rangle}\right)_i \langle 4 \rangle_i}{\sum_{{\rm events}}\left( W_{\langle 4 \rangle} \right)_i}
\end{split}
\end{equation}
while the weights are the number of two- and four-particle combinations:
\begin{equation}
W_{\langle 2 \rangle} \equiv M (M-1),
\end{equation}
\begin{equation}
W_{\langle 4 \rangle} \equiv M (M-1) (M-2) (M-3).
\end{equation}
Choosing the multiplicity weights above can make the final multi-particle azimuthal correlations free of
multiplicity fluctuations~\cite{Ante_thesis}.
However, one can also use unit weights treating events with different multiplicity equally.
The two- and four-particle cumulants without detector bias then can be formulated as:
\begin{equation}
c_n\{2\} = \llangle 2 \rrangle
\end{equation}
\begin{equation}
c_n\{4\} = \llangle 4 \rrangle - 2 \cdot \llangle 2 \rrangle^2
\end{equation}

\begin{figure*}[ht]
\vskip 0cm
\includegraphics[width=0.7\textwidth]{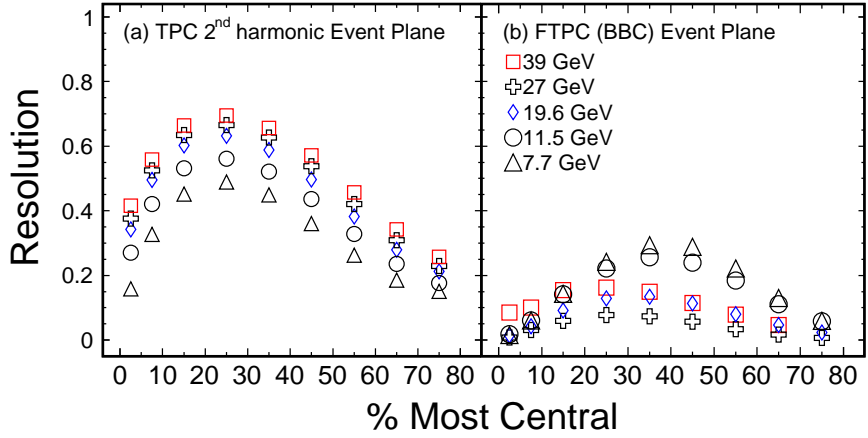}
\caption{(Color online) The event plane resolutions for \auau collisions at \sqrtsNN = 7.7, 11.5, 19.6, 27 and 39 GeV as a function of
collision centrality. Panel (a) shows the resolution of the second harmonic event plane from the TPC ($|\eta| < 1$). Panel (b) shows
the resolution for second harmonic event plane from the FTPCs ($2.5 < |\eta| < 4.0$) for 39 GeV and
second harmonic event plane resolution correction using the first-order event plane from the BBCs
($3.8 < |\eta| < 5.2$) for 7.7, 11.5, 19.6 and 27 GeV.
} \label{resolution}
\end{figure*}

The reference flow (e.g. integrated over $p_T$) can be estimated both from two- and four-particle cumulants:
\begin{equation}
v_n\{2\} = \sqrt{c_n\{2\}}
\end{equation}
\begin{equation}
v_n\{4\} = \sqrt[4]{-c_n\{4\}}
\end{equation}
Once the reference flow is estimated, we proceed to the calculation of differential flow (e.g. as a function of $p_T$)
of the particle of interest (POI)
which needs another two vectors $p$ and $q$. Particles used to estimate reference flow are called reference
particles (REP). For particles labeled as POI:
\begin{equation}
p_n \equiv \sum_{i=1}^{m_p} e^{i n \psi_i}.
\end{equation}
For particles labeled as both POI and REP:
\begin{equation}
q_n \equiv \sum_{i=1}^{m_p} e^{i n \psi_i}.
\end{equation}
Then the reduced single-event average two- and four-particle correlations are:
\begin{equation}
\langle 2^\prime \rangle = \frac{p_n Q^*_n - m_q}{m_p M - m_q}
\end{equation}
\begin{equation}
\begin{split}
\langle 4^\prime \rangle = & [p_n Q_n Q^*_n Q^*_n - q_{2n} Q^*_n Q^*_n - p_n Q_n Q^*_{2n} \\
						  & -2 \cdot M p_n Q^*_n - 2\cdot m_q |Q_n|^2 + 7 \cdot q_n Q^*_n \\
						  & -Q_n q^*_n + q_{2n} Q^*_{2n} + 2 \cdot p_n Q^*_n + 2 \cdot m_q M \\
						  & -6 \cdot m_q ] / [(m_p M - 3 m_q)(M-1)(M-2)]
\end{split}
\end{equation}
The event average can be obtained as follows:
\begin{equation}
\llangle 2^\prime \rrangle = \frac{\sum_{{\rm events}} (w_{\langle 2^\prime \rangle})_i \langle 2^\prime \rangle_i}{\sum_{i=1}^N (w_{\langle 2^\prime \rangle})_i}
\end{equation}
\begin{equation}
\llangle 4^\prime \rrangle = \frac{\sum_{{\rm events}} (w_{\langle 4^\prime \rangle})_i \langle 4^\prime \rangle_i}{\sum_{i=1}^N (w_{\langle 4^\prime \rangle})_i}
\end{equation}
Multiplicity weights are:
\begin{equation}
w_{\langle 2^\prime \rangle} \equiv m_p M - m_q
\end{equation}
\begin{equation}
w_{\langle 4^\prime \rangle} \equiv (m_p M - m_q)(M-1)(M-2)
\end{equation}
The two- and four-particle differential cumulants without detector bias are given by:
\begin{equation}
d_n\{2\} = \llangle 2^\prime \rrangle
\end{equation}
\begin{equation}
d_n\{4\} = \llangle 4^\prime \rrangle - 2\cdot \llangle 2^\prime \rrangle \llangle 2 \rrangle
\end{equation}
Equations for the case of detectors without uniform acceptance can be found in Ref.~\cite{Bilandzic:2010jr}.
Estimations of differential flow are expressed as:
\begin{equation}
v^\prime_n \{2\} = \frac{d_n\{2\}}{\sqrt{c_n\{2\}}}
\end{equation}
\begin{equation}
v^\prime_n \{4\} = \frac{d_n\{4\}}{-c_n\{2\}^{3/4}}
\end{equation}

\section{Results}
\label{sect_results}
\subsection{The event plane resolution}

\begin{figure*}[ht]
\vskip 0cm
\includegraphics[width=0.7\textwidth]{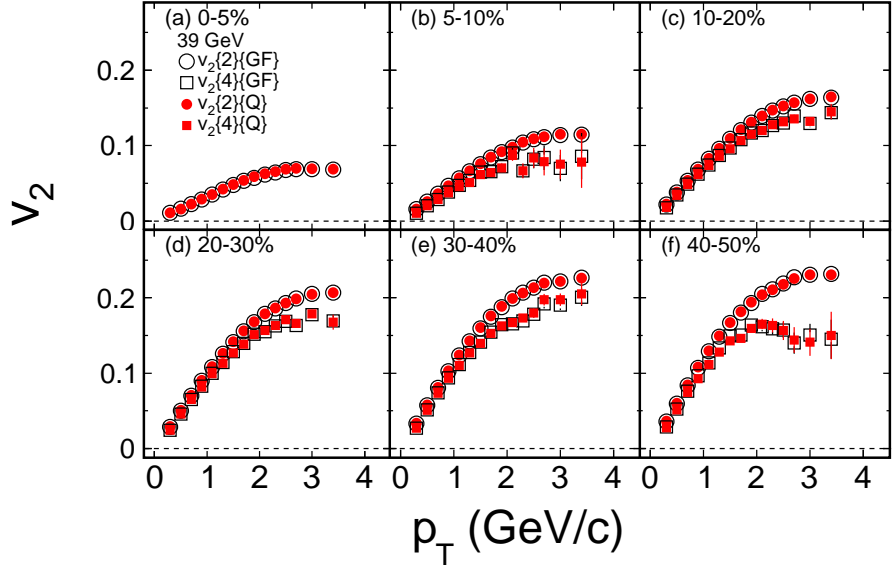}
\caption{(Color online) The comparison of $v_2$ as a function of $p_T$ between GF-cumulant (open symbols) and Q-cumulant (full symbols) methods in Au+Au collisions at \sqrtsNN = 39 GeV.
$v_{2}\{4\}$ fails in most central ($0 - 5\%$) collisions due to the small values of $v_2$ and large $v_2$ flucuations.
} \label{cumu_com}
\end{figure*}

\begin{figure*}[ht]
\vskip 0cm
\includegraphics[width=1\textwidth]{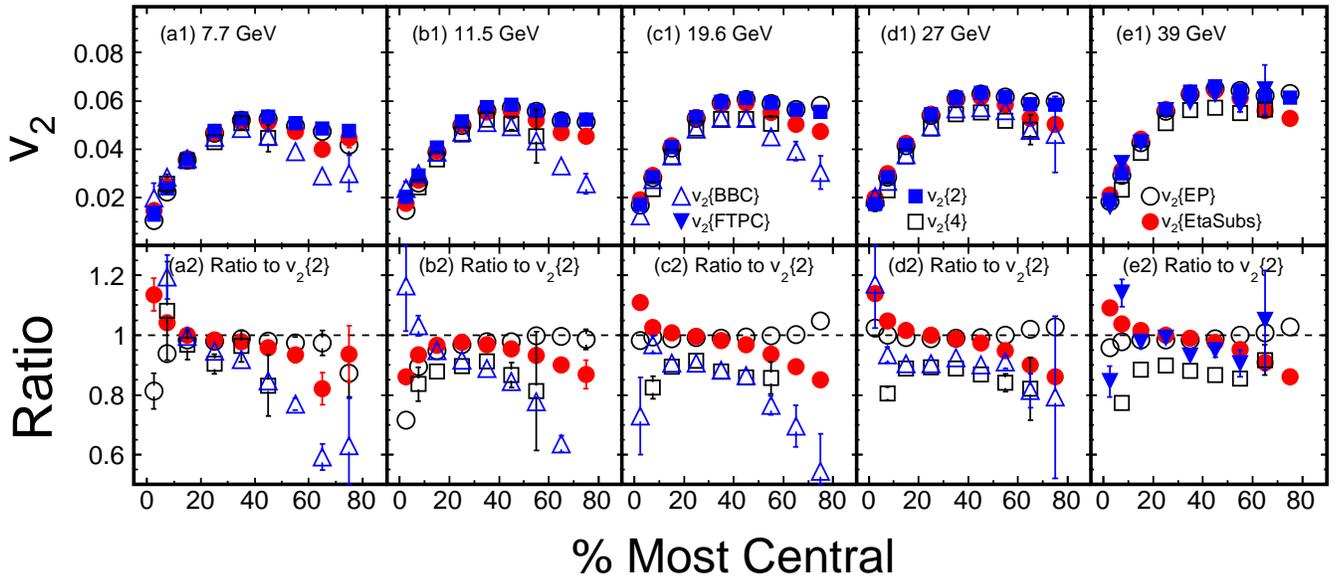}
\caption{(Color online)
The $p_T$ ($> 0.2$ GeV/$c$) and $\eta$ ($|\eta| < 1$) integrated $v_2$ as a function of collision centrality
for \auau collisions at \sqrtsNN = 7.7 GeV (a1), 11.5 GeV (b1), 19.6 GeV (c1) , 27 GeV (d1) and 39 GeV (e1).
The results in the top panels are presented for several methods of obtaining $v_2$.
The bottom panels show the ratio of $v_2$ obtained using the various techniques, with respect to $v_{2}\{2\}$.
The error bars shown are statistical.
} \label{integral_v2}
\end{figure*}

\begin{figure*}[ht]
\vskip 0cm
\includegraphics[width=1\textwidth]{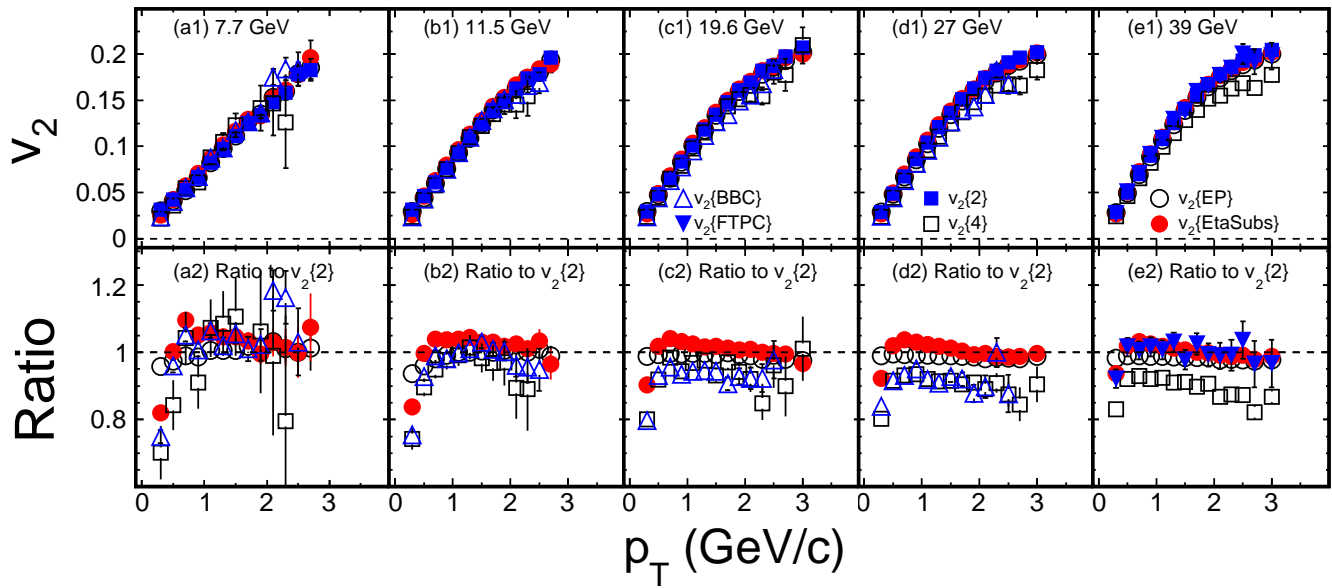}
\caption{(Color online)
The $v_2$ as a function of $p_T$ for $20 - 30\%$ central \auau collisions at midrapidity for \sqrtsNN =
7.7 GeV (a1), 11.5 GeV (b1), 19.6 GeV (c1) , 27 GeV (d1) and 39 GeV (e1). The top panels show $v_2$ vs. $p_T$
using various methods as labeled in the figure and discussed in the text. The bottom panels show the ratio of
$v_2$ measured using the various methods with respect to $v_2\{2\}$.
} \label{v2_cent_2030}
\end{figure*}

To investigate the non-flow correlations and $v_2$ fluctuations of the \vtwo measurements, the event planes from different detectors
and the cumulant method are used in the analysis.
The event planes are determined from the TPC in the midrapidity region, and the FTPC/BBC at forward rapidity.
The $\eta$ gap between FTPC/BBC to TPC could reduce the non-flow contribution in the \vtwo measurement~\cite{cucuv2STAR}.
Figure~\ref{resolution} shows the event plane resolution from TPC (panel (a)) and BBC (FTPC) (panel (b)).
The resolution of the TPC second harmonic event plane increases as the collision energy increases, as the resolution depends
on the multiplicity and the \vtwo signal~\cite{v2Methods}. Due to limited statistics, the FTPC event plane is used only
for the 39 GeV data set where the BBC event plane cannot be used because of the poor resolution.
The resolution of the FTPC event plane is about four times lowers than the TPC event plane.  The BBC is used to determine
the event plane for the 7.7, 11.5, 19.6 and 27 GeV data sets. Note the BBC event plane is based on the first harmonic,
as the $v_1$ signal is significant in the rapidity region covered by the BBC.
The qualitively different centrality dependence of the FTPC and BBC event plane resolutions is because of the different centrality dependence of $v_1$ and $v_2$.

\subsection{Method comparison}
\begin{figure*}[ht]
\vskip 0cm
\includegraphics[width=0.7\textwidth]{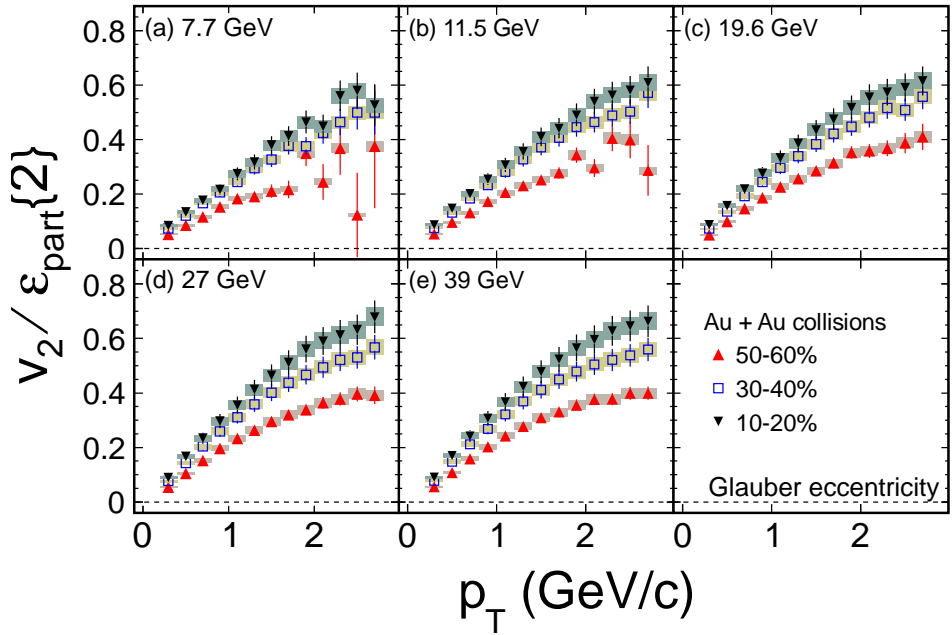}
\caption{(Color online)
The \vtwo over $\varepsilon$ (Glauber) as a function of \pt for various collision centralities ($10 - 20\%$, $30 - 40\%$ and $50 - 60\%$)
in Au + Au collisions at midrapidity. Panel (a), (b), (c), (d) and (e) show the results for \sqrtsNN = 7.7, 11.5, 19.6, 27 and 39 GeV respectively.
The data are from $v_2\{\rm {EtaSubs}\}$. The error bars and shaded boxes represent the statistical and systematic uncertainties respectively, as described in Sec. IV C.
} \label{v2_pT_galuber_all_Cent}
\end{figure*}

\begin{figure*}[ht]
\vskip 0cm
\includegraphics[width=0.7\textwidth]{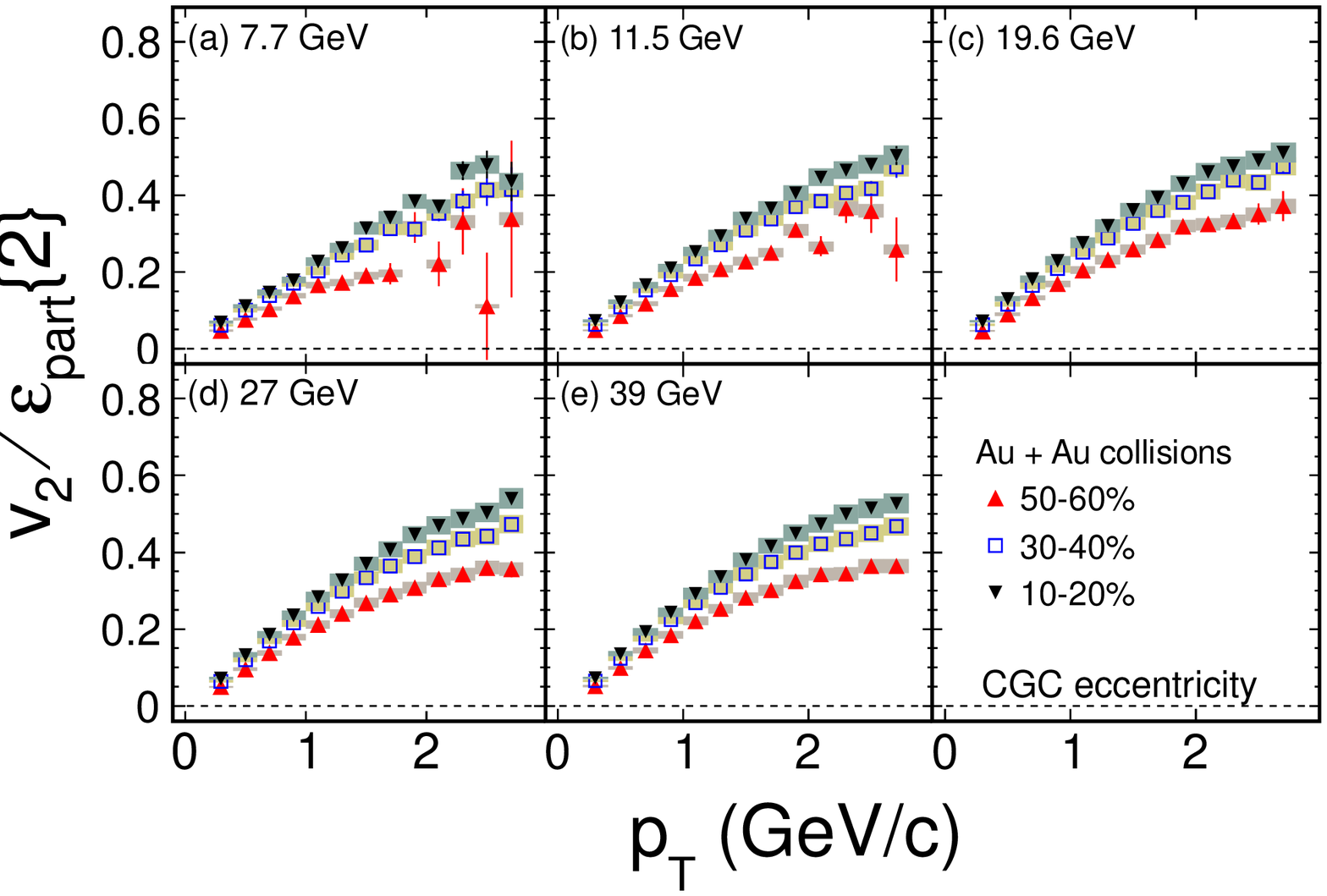}
\caption{(Color online)
The \vtwo over $\varepsilon$ (CGC) as a function of \pt for various collision centralities ($10 - 20\%$, $30 - 40\%$ and $50 - 60\%$)
in Au + Au collisions at midrapidity. Panel (a), (b), (c), (d) and (e) show the results for \sqrtsNN = 7.7, 11.5, 19.6, 27 and 39 GeV respectively.
The data are from $v_2\{\rm {EtaSubs}\}$. The error bars and shaded boxes represent the statistical and systematic uncertainties respectively, as described in Sec. IV C.
} \label{v2_pT_cgc_all_Cent}
\end{figure*}

The comparison of $v_2$ as a function of $p_T$ between the GF-cumulant and Q-cumulant methods is shown in Fig.~\ref{cumu_com} for six collision
centralities in Au+Au collisions at $\sqrt{s_{NN}}$ = 39 GeV.
The GF-cumulant and Q-cumulant methods agree within 5\% at all five collision energies.
Compared to GF-cumulant method, the recently developed Q-cumulant is the exact cumulant method~\cite{Bilandzic:2010jr}.
The observation of consistency between the two methods at BES energies implies the GF-cumulant is a good approximation.
The cumulant method (GF-cumulant or Q-cumulant) used in the
analysis does not cause difference in the comparison with other experimental results and theoretical calculations.
To be consistent with the previous STAR results, we will hereafter show only results from the GF-cumulant method.

Other method comparisons are shown in Figs.~\ref{integral_v2} and \ref{v2_cent_2030} for inclusive charged hadrons in \auau collisions at
\sqrtsNN = 7.7 GeV (a1), 11.5 GeV (b1), 19.6 GeV (c1), 27 GeV (d1) and 39 GeV (e1).
As the $v_2$ measurements from various methods are obtained using charged tracks recorded at midrapidity ($|\eta| < 1$), the statistical errors on the results from
the different $v_2$ methods are thus correlated. The conclusions on the differences in $v_{2}$ values from different methods are based on the systematic trends observed for the
corresponding ratios with respect to $v_{2}\{2\}$.
Figure~\ref{integral_v2} shows $v_2$
integrated over $0.2 < p_T < 2.0$ GeV/$c$ and $|\eta| < 1$ versus centrality. For comparison purposes, the integrated
\vtwo values for all methods are divided by the values of the 2-particle cumulant method ($v_2\{2\}$) and plotted in panels
(a2) through (e2). The results of the 4-particle cumulants are systematically lower than the other methods, except for $v_2\{\rm {FTPC/BBC}\}$.
The difference is about
$10-20\%$ in 39, 27 and 19.6 GeV, $10-15\%$ in 11.5 GeV and $5-10\%$ in 7.7 GeV.
The $\eta$ sub-event values for peripheral collisions ($50-60\%$ to $70-80\%$) drop below the 2-particle and TPC event plane results,
indicating the $\eta$ sub-event method could reduce some non-flow correlations for peripheral collisions.
Non-flow correlations are defined as correlations not related to the reaction plane.
The dominant non-flow correlations originating from two-particle correlations (such as HBT correlations, resonance decay) scale as $1/N$~\cite{v2Methods},
where $N$ is the multiplicity of particles
used to determine the event plane. Thus the non-flow contribution is larger in peripheral collisions. In mid-central and
peripheral collisions ($10-20\%$ to $40-50\%$), the data of $v_2\{\rm {BBC}\}$ from 7.7, 11.5, 19.6 and 27 GeV are consistent with
$v_2\{4\}$ and lower than other methods. It suggests the first-order (BBC) event plane suppresses the second-order non-flow and/or fluctuation effects.
Within statistical errors, the results of $v_2\{\rm {FTPC}\}$ from \auau collisions at \sqrtsNN = 39 GeV
are close to $v_2\{2\}$, $v_2\{\rm {EP}\}$ and $v_2\{\rm {EtaSubs}\}$ in semi-central collisions ($10-20\%$ to $20-30\%$).
In the peripheral collisions ($30-40\%$ to $60-70\%$), $v_2\{\rm {FTPC}\}$ falls between $v_2\{\rm {EtaSubs}\}$ and $v_2\{4\}$.
It indicates that the $\eta$ gap between TPC and FTPC reduces the non-flow contribution.

The $p_T$ differential $v_2$ from various methods for the $20-30\%$ centrality bin are shown in the upper panels of Fig.~\ref{v2_cent_2030}.
For comparison, the $v_2$ from other methods are divided by the results of the 2-particle cumulant method and shown in the lower panels
of Fig.~\ref{v2_cent_2030}. It can be seen that the difference of $v_2\{2\}$ compared to $v_2\{\rm {FTPC/BBC}\}$, $v_2\{2\}$ and $v_2\{\rm {EtaSubs}\}$ depends on the $p_T$ range.
A  larger difference can be observed in the low $p_T$ region ($p_T <$ 1 GeV/$c$). Beyond $p_T$ = 1 GeV/$c$ the difference stays
constant in the measured $p_T$ range. The difference between $v_2\{\rm {FTPC/BBC}\}$ and $v_2\{4\}$ is relatively small and less dependent on $p_T$.
It suggests the non-flow contribution to the event plane and 2-particle correlation methods depends on $p_T$.
Based on the interpretation in Ref.~\cite{flowreview1}, the difference between $v_2\{2\}^{2}$ and $v_2\{4\}^{2}$
is approximately equal to non-flow plus two times \vtwo fluctuations. The fact that the ratio of $v_2\{4\}$ to $v_2\{2\}$ is closer to
1 at the lower collision energies indicates the non-flow and/or \vtwo fluctuations in the \vtwo measurement depend on the collision energy.
One possible explanation is that the non-flow correlations from jets presumably decrease as the collision energy decreases.
The results of $v_2\{\rm {BBC}\}$ are found to be consistent with $v_2\{4\}$ in 7.7, 11.5, 19.6 and 27 GeV, while the $v_2\{\rm {FTPC}\}$ is larger than
$v_2\{4\}$ in 39 GeV. This consistency can be also observed in Fig.~\ref{integral_v2} for $10-20\%$ to $40-50\%$ centrality bins.
It indicates that the use of the first-order reaction plane (BBC event plane) to study the second harmonic flow eliminates flow fluctuations which are not correlated between different harmonics. The first-order BBC reaction plane is struck by nucleon spectators for these beam energies. The contribution of spectators makes the BBC event plane more sensitive to the reaction plane. This could partly explain the consistency between
$v_2\{\rm {BBC}\}$ and $v_2\{4\}$ mentioned above.
More studies of the collision energy dependence of non-flow and flow fluctuations will be discussed in another paper.

\begin{figure*}[ht]
\vskip 0cm
\includegraphics[width=0.7\textwidth]{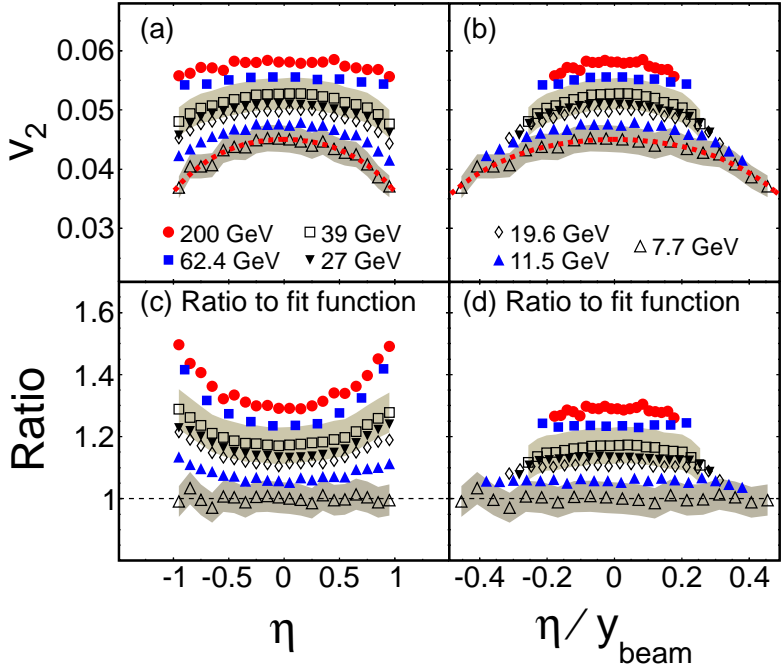}
\caption{(Color online)
Panel (a) shows the $v_2\{\rm {EP}\}$ vs. $\eta$  for $10 - 40\%$ centrality in Au + Au collisions at \sqrtsNN =
7.7, 11.5, 19.6, 27, 39, 62.4 and 200 GeV.
Panel (c) shows the ratio of $v_2$ vs. $\eta$ for all \sqrtsNN with respect to the fit curve.
Panel (b) shows the $v_2\{\rm {EP}\}$ vs. $\eta$/$y_{{\rm beam}}$.
Panel (d) shows the ratio of $v_2$ vs. $\eta$/$y_{{\rm beam}}$ for all \sqrtsNN with respect to the fit curve.
The data for \sqrtsNN = 62.4 and 200 GeV
are from refs.~\cite{star_v2cen, star_auau62.4, v2_4_200_62}.
The dashed red curves show the empirical fit to the results from Au + Au collisions at \sqrtsNN = 7.7 GeV.
The bands show the systematic uncertainties as described in Sec. IV C.
} \label{v2_eta_two}
\end{figure*}

\subsection{Systematic uncertainties}

Different $v_2$ methods show different sensitivities to non-flow correlations and $v_2$ fluctuations.
In previous STAR publications,
the differences between different methods were regarded as systematic uncertainties~\cite{star_fv2, star_v2cen}.
A great deal of progress has revealed that some of these differences are not due to systematic uncertainties in different methods,  but due to different sensitivities to non-flow and flow fluctuation effects~\cite{flowfluc, epart21}.
The four particle cumulant method is less sensitive to non-flow correlations~\cite{cumulant1, cumulant2} and has a negative contribution from flow fluctuations.
$v_2$ measurements from the two particle cumulant method and the event plane method (the second harmonic event plane) have positive contributions from flow fluctuations as well as non-flow.
It was also noticed that four particle cumulant results should be very close to flow in the reaction plane, while the two particle cumulant measures flow in the participant plane~\cite{flowfluc, epart21}.
Further, because of the large pseudorapidity gap between the BBC/FTPC and TPC, $v_{2}\{\rm BBC\}$ and $v_{2}\{\rm FTPC\}$ are most insensitive to non-flow correlations.

We estimate the systematic uncertainty on event plane flattening methods for $v_2\{\rm {EP}\}$ and $v_2\{\rm {EtaSubs}\}$ by the difference between them and find it to be negligible (below $1\%$).
A $5\%$ systematic uncertainty on $v_{2}\{\rm BBC\}$, $v_{2}\{\rm FTPC\}$, $v_2\{\rm {EP}\}$ and $v_2\{\rm {EtaSubs}\}$
is estimated by varying cut parameters (e.g. collision vertex position, the distance of closest approach to the primary vertex for the tracks, and the number of fit points used for reconstruction of the tracks).
The systematic uncertainties on $v_{2}\{2\}$ and $v_{2}\{4\}$ are based on the difference between Q-cumulant and GF-cumulant methods (5\%) as well as cut variations (5\%).
All the percentage uncertainties are relative to the $v_2$ value.

\begin{figure*}[ht]
\vskip 0cm
\includegraphics[width=0.7\textwidth]{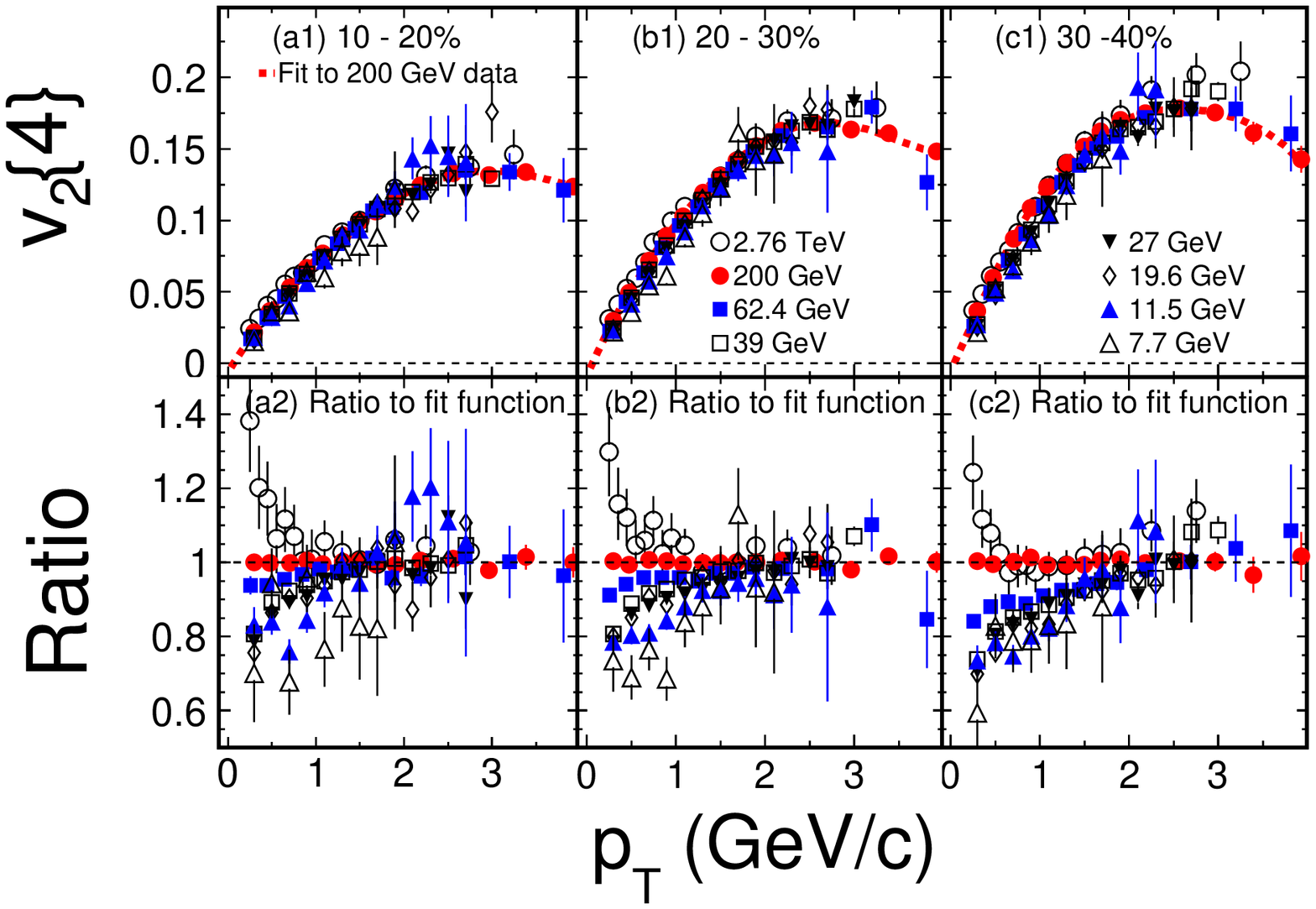}
\caption{(Color online)
The top panels show $v_2\{4\}$ vs. $p_T$ at midrapidity for various collision energies (\sqrtsNN = 7.7 GeV to 2.76 TeV).
The results for \sqrtsNN = 7.7 to 200 GeV are for \auau collisions and those for 2.76 TeV are for Pb + Pb collisions.
The dashed red curves show the empirical fits to the results from \auau collisions at \sqrtsNN = 200 GeV. The bottom panels show
the ratio of $v_2\{4\}$ vs. $p_T$ for all \sqrtsNN with respect to the fit curve. The results are shown for
three collision centrality classes: $10 - 20\%$ (a1), $20 - 30\%$ (b1) and $30 - 40\%$ (c1).
Error bars are shown only for the statistical uncertainties.}
\label{v2_4_pt_beam_energy}
\end{figure*}

\begin{figure*}[ht]
\vskip 0cm
\includegraphics[width=0.7\textwidth]{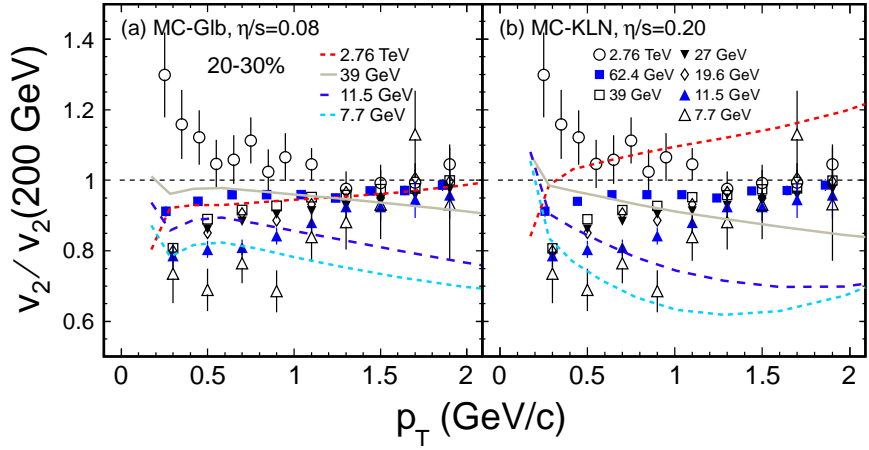}
\caption{(Color online) The experimental data (symbols) are the same as in Fig.~\ref{v2_4_pt_beam_energy}~(b2).
The lines represent the viscous hydrodynamic calculations from Ref.~\cite{hydroBES} based on (a) MC-Glauber initial conditions and
$\eta/s$ = 0.08 (b) MC-KLN initial conditions and $\eta/s$ = 0.20.}

\label{compare_hydro}
\end{figure*}

\begin{figure*}[ht]
\vskip 0cm
\includegraphics[width=0.7\textwidth]{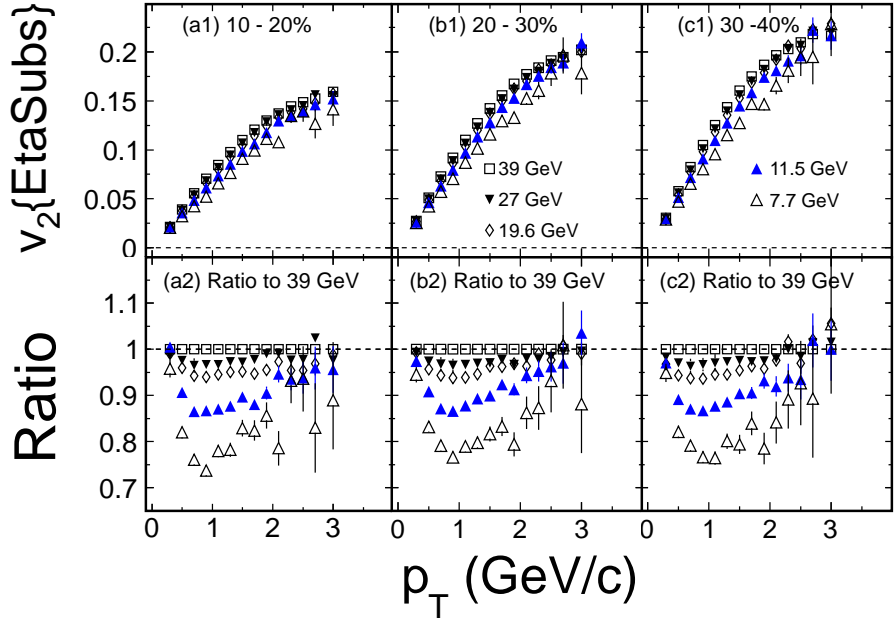}
\caption{(Color online)
The top panels show $v_{2}\{\rm EtaSubs\}$ vs. $p_T$ at midrapidity for various collision energies (\sqrtsNN = 7.7 GeV to 39 GeV).
The bottom panels show
the ratio of $v_{2}\{\rm EtaSubs\}$ vs. $p_T$ for all \sqrtsNN with respect to the 39 GeV data. The results are shown for
three collision centrality classes: $10 - 20\%$ (a1), $20 - 30\%$ (b1) and $30 - 40\%$ (c1).
Error bars are shown only for the statistical uncertainties.
} \label{v2_etasub}
\end{figure*}

\begin{figure*}[ht]
\vskip 0cm
\includegraphics[width=0.9\textwidth]{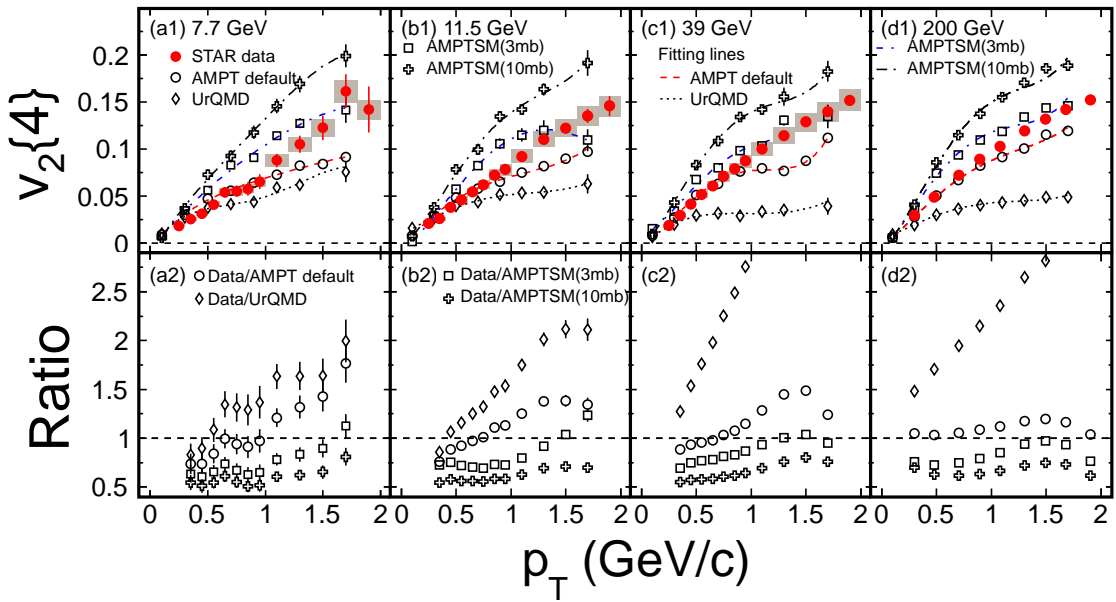}
\caption{(Color online)
The $v_2\{4\}$ as a function of $p_T$ for $20 - 30\%$ \auau collisions at \sqrtsNN = 7.7, 11.5, 39 and 200 GeV compared to corresponding results
from UrQMD, AMPT default version, and AMPT with string melting version (3 and 10 mb). The shaded boxes show the systematic uncertainties for the
experimental data of 7.7, 11.5 and 39 GeV.
The bottom panels show the ratio of data to the fit results of
the models.
} \label{v2_pT_model_com}
\end{figure*}

\section{Discussion}
\label{sect_discuss}
\subsection{Transverse momentum and centrality dependence of $v_2$}

\begin{table*}[ht]
\centering
\caption{The  \eparttwos~and transverse area $\left<S_{\rm part}\right>$ from the
Color Glass Condensate (CGC) model~\cite{CGCa,CGCb,CGCc, CGCd} calculations in Au + Au collisions
at \sqrtsNN = 7.7, 11.5, 19.6, 27, 39, 62.4 and 200 GeV. The errors are systematic
uncertainties.}
\begin{tabular}{cccccccccc} \hline\hline
Centrality (\%)&$0-5\%$&$5-10\%$&$10-20\%$&$20-30\%$&$30-40\%$&$40-50\%$&$50-60\%$&$60-70\%$&$70-80\%$\\ \hline
\hline
\multicolumn{10}{c}{\auau at \sqrtsNN = 7.7 GeV} \\
\hline
\eparttwos&$0.104\pm0.005$&$0.19\pm0.01$&$0.29\pm0.01$&$0.39\pm0.02$&$0.47\pm0.02$&$0.54\pm0.03$&$0.59\pm0.03$&$0.62\pm0.03$&$0.51\pm0.02$\\
$\left<S_{\rm part}\right>$ (fm$^2$)&$25.9\pm1.3$&$21.8\pm1.1$&$17.5\pm0.9$&$13.4\pm0.7$&$10.2\pm0.5$&$7.7\pm0.4$&$5.5\pm0.3$&$3.6\pm0.2$&$1.8\pm0.1$\\
\hline
\multicolumn{10}{c}{\auau at \sqrtsNN = 11.5 GeV} \\
\hline
\eparttwos&$0.104\pm0.005$&$0.19\pm0.01$&$0.29\pm0.01$&$0.39\pm0.02$&$0.47\pm0.02$&$0.53\pm0.03$&$0.59\pm0.03$&$0.62\pm0.03$&$0.51\pm0.02$\\
$\left<S_{\rm part}\right>$ (fm$^2$)&$25.2\pm1.2$&$21.2\pm1.1$&$17.0\pm0.9$&$13.0\pm0.7$&$9.9\pm0.5$&$7.5\pm0.4$&$5.4\pm0.3$&$3.5\pm0.2$&$1.8\pm0.1$\\
\hline
\multicolumn{10}{c}{\auau at \sqrtsNN = 19.6 GeV} \\
\hline
\eparttwos&$0.105\pm0.005$&$0.19\pm0.01$&$0.29\pm0.01$&$0.39\pm0.02$&$0.47\pm0.02$&$0.53\pm0.03$&$0.58\pm0.03$&$0.61\pm0.03$&$0.51\pm0.02$\\
$\left<S_{\rm part}\right>$ (fm$^2$)&$24.4\pm1.2$&$20.6\pm1.0$&$16.6\pm0.9$&$12.6\pm0.7$&$9.7\pm0.5$&$7.3\pm0.4$&$5.3\pm0.3$&$3.5\pm0.2$&$1.8\pm0.1$\\
\hline
\multicolumn{10}{c}{\auau at \sqrtsNN = 27 GeV} \\
\hline
\eparttwos&$0.105\pm0.005$&$0.19\pm0.01$&$0.29\pm0.01$&$0.39\pm0.02$&$0.47\pm0.02$&$0.53\pm0.03$&$0.58\pm0.03$&$0.61\pm0.03$&$0.51\pm0.02$\\
$\left<S_{\rm part}\right>$ (fm$^2$)&$24.1\pm1.2$&$20.3\pm1.0$&$16.4\pm0.8$&$12.5\pm0.6$&$9.6\pm0.5$&$7.2\pm0.4$&$5.3\pm0.3$&$3.5\pm0.2$&$1.8\pm0.1$\\
\hline
\multicolumn{10}{c}{\auau at \sqrtsNN = 39 GeV} \\
\hline
\eparttwos&$0.105\pm0.005$&$0.19\pm0.01$&$0.29\pm0.01$&$0.39\pm0.02$&$0.47\pm0.02$&$0.53\pm0.03$&$0.58\pm0.03$&$0.61\pm0.03$&$0.50\pm0.02$\\
$\left<S_{\rm part}\right>$ (fm$^2$)&$23.9\pm1.2$&$20.1\pm1.0$&$16.2\pm0.8$&$12.4\pm0.6$&$9.5\pm0.5$&$7.2\pm0.4$&$5.3\pm0.3$&$3.5\pm0.2$&$1.8\pm0.1$\\
\hline
\multicolumn{10}{c}{\auau at \sqrtsNN = 62.4 GeV} \\
\hline
\eparttwos&$0.105\pm0.005$&$0.19\pm0.01$&$0.29\pm0.01$&$0.39\pm0.02$&$0.47\pm0.02$&$0.53\pm0.03$&$0.58\pm0.03$&$0.61\pm0.03$&$0.50\pm0.02$\\
$\left<S_{\rm part}\right>$ (fm$^2$)&$23.7\pm1.2$&$20.0\pm1.0$&$16.1\pm0.8$&$12.3\pm0.6$&$9.4\pm0.5$&$7.2\pm0.4$&$5.3\pm0.3$&$3.5\pm0.2$&$1.8\pm0.1$\\
\hline
\multicolumn{10}{c}{\auau at \sqrtsNN = 200 GeV} \\
\hline
\eparttwos&$0.104\pm0.005$&$0.19\pm0.01$&$0.29\pm0.01$&$0.39\pm0.02$&$0.47\pm0.02$&$0.53\pm0.03$&$0.57\pm0.03$&$0.60\pm0.03$&$0.49\pm0.02$\\
$\left<S_{\rm part}\right>$ (fm$^2$)&$23.7\pm1.2$&$20.0\pm1.0$&$16.1\pm0.8$&$12.3\pm0.6$&$9.4\pm0.5$&$7.2\pm0.4$&$5.3\pm0.3$&$3.6\pm0.2$&$1.9\pm0.1$\\
      \hline
      \hline
\end{tabular}

\label{tab:CGC}
\end{table*}

The centrality dependence of $p_T$ differential \vtwo with respect to the initial eccentricity has been studied in detail for \auau and \cucu collisions in
\sqrtsNN = 200 and 62.4 GeV~\cite{star_v2cen, cucuv2STAR}. The larger magnitude of $v_2$ in the more peripheral collisions
could be due to the larger initial eccentricity in coordinate space for the more peripheral collisions. The participant eccentricity
is the initial configuration space eccentricity of the participant nucleons defined by Eq.~(3).  The root-mean-square participant
eccentricity, $\varepsilon_{\rm part}\{2\}$, is calculated from the Monte Carlo Glauber model~\cite{ms03,glauber}  (Tab.~\ref{tab:glauber}) and Color Glass Condensate (CGC) model~\cite{CGCa,CGCb,CGCc, CGCd} (Tab.~\ref{tab:CGC}).
The event plane is constructed from hadrons which have their origin in participant nucleons.
At the same time, the event plane resolution ($\eta$ sub-event) is less than 0.5. Thus, what we actually measure is the root-mean-square of $v_{2}$ with
respect to the participant plane~\cite{epart21}. In this case, \eparttwos~is the appropriate measure of the initial geometric anisotropy taking
the event-by-event fluctuations into account~\cite{eparttwo, epart2, epart21}. In Figs.~\ref{v2_pT_galuber_all_Cent} and \ref{v2_pT_cgc_all_Cent},
the centrality dependence of $p_T$ differential $v_2$ over eccentricity is shown for \auau collisions at \sqrtsNN = 7.7, 11.5, 19.6, 27 and 39 GeV.
For all five collision energies, the centrality dependence of $v_2$($p_T$) is observed to be similar to that at higher collision energies (62.4 and 200 GeV)
of \auau and \cucu colliding systems. That central collisions in general have higher $v_2/\varepsilon$ than peripheral collisions
is consistent with the picture that collective interactions are stronger in collisions with larger numbers of participants.

\subsection{Pseudorapidity dependence}
The panel (a) of Fig.~\ref{v2_eta_two} shows $v_2$ as a function of pseudorapidity for Au + Au collisions at \sqrtsNN =
7.7, 11.5, 19.6, 27, 39, 62.4 and 200 GeV in mid-central ($10 - 40\%$) collisions. The data for \sqrtsNN = 62.4 and 200 GeV
are from refs.~\cite{star_v2cen, star_auau62.4, v2_4_200_62}.
To facilitate comparison with 62.4 and 200 GeV data, the results of $v_{2}\{\rm EP\}$ are selected for the rest of the
collision energies. The 7.7 GeV data are empirically fit by the following function:
\begin{equation}
v_2(\eta) = p_0 + p_{1}\eta^2 + p_{2}\eta^4
\label{v2_eta_fit}
\end{equation}
with parameters $p_0 = 0.0450 \pm 0.0002$, $p_1 = -0.0064 \pm 0.0015$, $p_2 = -0.0024 \pm 0.0017$.
For clarity, the panel (c) of Fig.~\ref{v2_eta_two} shows the ratio of $v_2$($\eta$) with respect to this fit function.
The pseudorapidity dependence of $v_2$ indicates a change in shape as we move from \sqrtsNN = 200 GeV to 7.7 GeV within
our measured range $-1 <\eta < 1$.

To investigate the collision energy dependence of  the $v_2$($\eta$) shape, in panel (b) and (d) of Fig.~\ref{v2_eta_two},
the same $v_2$ results have been plotted as a function of pseudorapidity divided by beam rapidity. The data
of 7.7 GeV are fit by Eq.~(\ref{v2_eta_fit}) with parameters
$p_0 = 0.0450 \pm 0.0002$, $p_1 = -0.0279 \pm 0.0064$, $p_2 = -0.0464 \pm 0.0325$. The beam rapidities are 2.09, 2.50, 3.04, 3.36, 3.73, 4.20 and 5.36
for \sqrtsNN = 7.7, 11.5, 19.6, 27, 39, 62.4 and 200 GeV respectively.
After dividing pseudorapidity by the beam rapidity, the shape of $v_2$ seems similar at all collision energies.
The approximate beam rapidity scaling on the $v_{2}(\eta)$ shape suggests the change in shape may be
related to the final particle density. Higher particle density indicates higher probability of interaction which can generate larger collective flow.

\subsection{Energy dependence}
One of the most important experimental observations at RHIC is the significant $v_2$ signal in the top energy of \auau
collisions~\cite{star_130v2, runII200gevV2} (more than $50\%$ larger than at the SPS~\cite{SPSv2}). It could be interpreted
as the observation of a higher degree of thermalization than at lower collision energies~\cite{star_130v2}. The BES data
from the RHIC-STAR experiment offers an opportunity to study the collision energy dependence of $v_2$ using a wide acceptance
detector at midrapidity. Figure~\ref{v2_4_pt_beam_energy} shows the $p_T$ dependence of $v_2\{4\}$ from
\sqrtsNN = 7.7 GeV to 2.76 TeV in $10 - 20\%$ (a1), $20 - 30\%$ (b1) and $30 - 40\%$ (c1) centrality bins, where the
ALICE results in Pb + Pb collisions at $\sqrt{s_{NN}}$ = 2.76 TeV are taken from Ref.~\cite{alicev2}. The reasons to select
the results of $v_2\{4\}$ for the comparison are the following: 1) keep the method for $v_2$ measurements consistent
with the published results of ALICE; 2) $v_2\{4\}$ is insensitive to non-flow correlations.
The 200 GeV data is empirically fit by a fifth order polynomial function. The parameters for the fit function are listed
in Table~\ref{tab:parameters}.
\begin{table*}[ht]
\caption{\label{tab:parameters}
Summary of the parameters for the fit functions to the results of $v_2\{4\}$ vs. $p_T$ in Au+Au collisions at \sqrtsNN = 200 GeV.
}
\centering
\begin{tabular}{|c||c|c|c|c|c|c|c|c|c|} \hline
     Parameters   & \ $p_0$ & \ $p_1$ & \ $p_2$   & \ $p_3$ & \ $p_4$ & \ $p_5$  \\ \hline \hline
     $10 - 20\%$& $-0.00730 \pm 0.00114$& $0.10785 \pm 0.00598$& $-0.03941 \pm 0.01038$& $0.01508 \pm 0.00767$& $-0.00411 \pm 0.00246$& $0.00041 \pm 0.00028$ \\ \hline
     $20 - 30\%$& $-0.00890 \pm 0.00096$& $0.14250 \pm 0.00500$& $-0.05206 \pm 0.00869$& $0.02156 \pm 0.00642$& $-0.00685 \pm 0.00206$& $0.00077 \pm 0.00023$ \\ \hline
     $30 - 40\%$& $-0.00581 \pm 0.00206$& $0.14526 \pm 0.01089$& $-0.00529 \pm 0.01910$& $-0.02409\pm 0.01419$&  $0.00797 \pm 0.00456$& $-0.00084\pm 0.00052$ \\ \hline
\end{tabular}
\end{table*}
For comparison, the $v_2$ from other energies are divided by the fit and shown in the lower panels of
Fig.~\ref{v2_4_pt_beam_energy}. We choose 200 GeV data as the reference because the statistical errors are smallest.
For $p_T$ below 2~\GeVc, the $v_2$ values rise with increasing collision energy. Beyond $p_T = 2~\GeVc$
the $v_2$ results show comparable values within statistical errors.

The increase of $v_{2}(p_{T})$ as a function of energy could be
due to the change of chemical composition from low to high energies~\cite{star_ub} and/or larger collectivity at the higher collision energy.
The baryonic chemical potential varies a lot (20 - 400 MeV) from
200 to 7.7 GeV~\cite{star_ub}. The baryon over meson ratio is larger in lower
collisions energies. The difference of $v_2$ for baryon and meson,
for example proton $v_2$ $<$ pion $v_2$ for $p_T$ below 2 GeV/$c$, could partly explain
the collision energy dependence.
Further, in Fig.~\ref{compare_hydro} we compare the experimental data from Fig.~\ref{v2_4_pt_beam_energy}~(b2) to the viscous hydrodynamic calculations~\cite{hydroBES}.
As the collision energy varies from \sqrtsNN = 7.7 to 2760 GeV, the experimental data show larger splitting in the lower $p_T$ region
and converge at the intermediate range ($p_T\sim$ 2 GeV/$c$);
while, in the pure viscous hydrodynamic simulations, the splitting increases with $p_T$.
The $p_T$ dependence of the $v_2$ ratio cannot be reproduced by pure viscous hydrodynamic simulations with a constant shear viscosity to
entropy density ratio ($\eta/s$), and zero net baryon density.
The comparison suggests that a quantitative study at lower collision energies requires a more serious theoretical approach,
like 3D viscous hydro + UrQMD with a consistent equation of state at non-zero baryon chemical potential.

Figure~\ref{v2_etasub} shows the energy dependence of $v_{2}\{\rm EtaSubs\}$. Larger $v_{2}\{\rm EtaSubs\}$ values are
observed at higher collision energy for a selected $p_T$ bin, but the $p_T$ dependence of the difference is quite different
from $v_2\{4\}$. The ratios to 39 GeV data for each collision energy first decrease as a function of $p_T$, then slightly increase in the $p_T$ region of
1 - 2.5 GeV/$c$. The different trend of the energy dependence of $v_2$ from $v_2\{4\}$ and $v_{2}\{\rm EtaSubs\}$ is interpreted
as due to the different sensitivity of the $v_2$ methods to non-flow and/or flow fluctuations.

\subsection{Model comparisons}
To investigate the partonic and hadronic contribution to the final $v_2$ results from different collision energies, transport model calculations from
AMPT default (ver. 1.11), AMPT string-melting (ver. 2.11)~\cite{ampt} and UrQMD (ver 2.3)~\cite{urqmd} are compared with the new data presented.
The initial-parameter settings for the models follow the recommendation in the cited references.
The AMPT default and UrQMD models only take the hadronic
interactions into consideration, while the AMPT string-melting version incorporates both partonic and hadronic interactions.
The larger the parton cross section, the later the hadron cascade starts.

Figure~\ref{v2_pT_model_com} shows the comparison of $p_T$
differential $v_2\{4\}$ between model and data in the $20 - 30\%$ centrality bin.
The 200 GeV data are taken from~\cite{v2_4_200_62}.
The figure shows that UrQMD  underpredicts the measurements at
$\sqrt{s_{NN}}$ = 39 and 200 GeV in the $p_T$  range studied. The
differences are reduced as the collision energy decreases. That the ratio of
data to UrQMD results are closer to 1 at the lower collision energy
indicates that the contribution of hadronic interactions becomes more
significant at lower collision energies. The AMPT model with default
settings underpredicts the 200 GeV data, while the ratios of data to AMPT
default results show no significant change from 7.7 to 39 GeV.
The inconsistency between AMPT default and UrQMD makes the conclusion
model dependent.
The AMPT model with string-melting version with 3 and 10 mb parton cross
sections overpredicts the results at all collision energies from 7.7 to 200 GeV.
A larger parton cross section means stronger partonic interactions which
translate into a larger magnitude of $v_2$. The difference between
data and these AMPT model calculations seems to show no significantly
systematic change vs. collision energies. However, a recent study with
the AMPT model suggests hadronic potentials affect the final $v_2$ results
significantly when the collision energy is less than $\sqrt{s_{NN}}$ =
39 GeV~\cite{hadronic_potential}.

\section{Summary}
\label{sect_summary}
We have presented elliptic flow, $v_2$, measurements from \auau collisions at \sqrtsNN = 7.7, 11.5, 19.6, 27 and 39 GeV for inclusive charged
hadrons at midrapidity. To investigate non-flow correlations and $v_2$ fluctuations, various measurement methods have been used
in the analysis. The difference between $v_2\{2\}$ and $v_2\{4\}$ decreases with decrease in collision energy, indicating that non-flow contribution
and/or flow fluctuations decrease with a decrease in collision energy. The centrality and $p_T$ dependence of $v_2$ are similar to that observed at
higher RHIC collision energies. A larger $v_2$ is observed in more peripheral collisions. The pseudorapidity dependence of $v_2$ indicates
a change in shape from 200 GeV to 7.7 GeV within the measured range $-1 < \eta < 1$, but the results of $v_2$ versus pseudorapidity scaled by beam rapidity shows a similar trend for all collision energies. The comparison with \auau collisions at higher energies at RHIC (\sqrtsNN = 62.4 and 200 GeV)
and at LHC (Pb + Pb collisions at \sqrtsNN = 2.76 TeV) shows the $v_2\{4\}$ values at low $p_T$ ($p_T <$ 2.0 GeV/$c$) increase with increase in
collision energy implying an increase of collectivity.
The current viscous hydrodynamic simulations cannot reproduce the trend of the energy dependence of $v_{2}(p_{T})$.

The agreement between the data and UrQMD, which is based on hadronic
rescatterings, improves at lower collision energies, consistent with an
increasing role of the hadronic stage at these energies.
The inconsistency between AMPT default and UrQMD makes the conclusion
model dependent.
The comparison to AMPT model calculations seems to show no significantly systematic
change vs. collision energy, but improved calculations including
harmonic potentials may change the $v_2$ values from AMPT models at
lower collision energies.

These results set the baseline to study the number of constituent quark scaling of identified hadron $v_2$.
It also sets the stage for understanding the collision energy dependence of $v_2$ in the regime where the relative contribution of baryon and
mesons vary significantly.

\section{Acknowledgments}
We thank the RHIC Operations Group and RCF at BNL, the NERSC Center at LBNL and the Open Science Grid consortium for providing resources and support. This work was supported in part by the Offices of NP and HEP within the U.S. DOE Office of Science, the U.S. NSF, the Sloan Foundation, CNRS/IN2P3, FAPESP CNPq of Brazil, Ministry of Ed. and Sci. of the Russian Federation, NNSFC, CAS, MoST, and MoE of China, GA and MSMT of the Czech Republic, FOM and NWO of the Netherlands, DAE, DST, and CSIR of India, Polish Ministry of Sci. and Higher Ed., Korea Research Foundation, Ministry of Sci., Ed. and Sports of the Rep. of Croatia, and RosAtom of Russia.

%

%

\begin{thebibliography}{99}

\bibitem{flowreview1} S. A. Voloshin, A. M. Poskanzer and R. Snellings, arXiv:0809.2949 (2008).
\bibitem{flowreview2} P. Sorensen, arXiv:0905.0174 (2009).
\bibitem{flowreview3} R. Snellings, New J. Phys. {\bf 13}, 055008 (2011).



\bibitem{firsthydro} J. Y. Ollitrault, Phys. Rev. D {\bf 46}, 229 (1992).
\bibitem{hydroBES} C. Shen and U. Heinz, Phys. Rev. C {\bf 85}, 054902 (2012).
\bibitem{star_130v2} K. H. Ackermann {\it et al.} (STAR Collaboration), Phys. Rev. Lett. {\bf 86}, 402 (2001).
\bibitem{starklv2} J. Adams {\it et al.} (STAR Collaboration), Phys. Rev. Lett. {\bf 92}, 052302 (2004).
\bibitem{msv2} J. Adams {\it et al.} (STAR Collaboration), Phys. Rev. Lett. {\bf 95}, 122301 (2005).
\bibitem{starwp} J. Adams {\it et al.}, (STAR Collaboration), Nucl. Phys. A {\bf 757}, 102 (2005).
\bibitem{runII200gevV2} J. Adams {\it et al.} (STAR Collaboration), Phys. Rev. C {\bf 72}, 014904 (2005).
\bibitem{star_fv2} B. I. Abelev {\it et al.} (STAR Collaboration), Phys. Rev. Lett. {\bf 99}, 112301 (2007).
\bibitem{star_v2cen} B. I. Abelev {\it et al.} (STAR Collaboration), Phys. Rev. C {\bf 77}, 054901 (2008).
\bibitem{cucuv2STAR} B. I. Abelev {\it et al.} (STAR Collaboration), Phys. Rev. C {\bf 81}, 044902 (2010).
\bibitem{phenixv2_1} S. S. Adler {\it et al.} (PHENIX Collaboration), Phys. Rev. Lett {\bf 91}, 182301 (2003).
\bibitem{phenixprl} S. S. Adler {\it et al.} (PHENIX Collaboration), Phys. Rev. Lett {\bf 94}, 232302 (2005).
\bibitem{phobosv2} B. Alver {\it et al.} (PHOBOS Collaboration), Phys. Rev. Lett {\bf 98}, 242302 (2007).
\bibitem{na49v2} H. Appelshauser {\it et al.} (NA49 Collaboration), Phys. Rev. Lett {\bf 80}, 4136 (998).
\bibitem{alicev2} K. Aamodt {\it et al.} (ALICE Collaboration), Phys. Rev. Lett. {\bf 105}, 252302 (2010).
\bibitem{atlasv2} G. Aad {\it et al.} (ATLAS Collaboration), Phys. Lett. B {\bf 707}, 330 (2012).

\bibitem{bes}B. Mohanty (For STAR Collaboration), J. Phys. G {\bf 38}, 124023 (2011).
\bibitem{stat_model}
J. Cleymans {\it et al.}, Phys. Rev. C {\bf 73}, 034905 (2006);
F. Becattini, J. Manninen and M. Gazdzicki, Phys. Rev. C {\bf 73}, 044905 (2006);
A.Andronic, P.Braun-Munzinger and J. Stachel, Nucl. Phys. A {\bf 772}, 167 (2006).
\bibitem{star_ub} L. Kumar (For STAR Collaboration), J. Phys. G {\bf 38}, 124145 (2011).
\bibitem{chemfo} J. Cleymans {\it et al.}, Phys. Rev. C {\bf 73}, 034905 (2006).

\bibitem{lattice} Y. Aoki {\it et al.}, Nature {\bf 443}, 675 (2006).
\bibitem{order1}S. Ejiri, Phys. Rev. D {\bf 78}, 074507 (2008).
\bibitem{order2}M. A. Stephanov, Int. J. Mod. Phys. A {\bf 20}, 4387 (2005).
\bibitem{order}B. Mohanty, Nucl. Phys. A {\bf 830}, 899c (2009) and references therein.
\bibitem{CP_predict} P.F. Kolb, J. Sollfrank, U. Heinz, Phys. Rev. C {\bf 62} (2000) 054909.
\bibitem{sorge}H. Sorge, Phys. Rev. Lett. {\bf 82}, 2048 (1999).
\bibitem{v2Methods} A. M. Poskanzer and S. A. Voloshin, Phys. Rev. C \textbf{58} 1671 (1998).
\bibitem{2part}S. Wang et al., Phys. Rev. C {\bf 44}, 1091 (1991).
\bibitem{cumulant1}N. Borghini, P. M. Dinh, and J.-Y. Ollitrault, Phys. Rev. C \textbf{63}, 054906 (2001).
\bibitem{cumulant2}N. Borghini, P. M. Dinh, and J.-Y. Ollitrault, Phys. Rev. C \textbf{64}, 054901 (2001).
\bibitem{LYZ} R. S. Bhalerao, N. Borghini and J. Y. Ollitrault, Nucl. Phys. A {\bf 727}, 373 (2003).
\bibitem{flowfluc}S. A. Voloshin, A. M. Poskanzer, A. Tang and G. Wang, Phys. Lett. B {\bf 659}, 537 (2008).
\bibitem{nasim}Md. Nasim et. al., Phys. Rev. C {\bf 82}, 054908 (2010).


\bibitem{BBC}C. A. Whitten Jr. (For STAR Collaboration), AIP Conf. Proc. {\bf 980}, 390 (2008).
\bibitem{ZDC} C. Adler {\it et al.} (STAR Collaboration), Nucl. Instrum. Methods A {\bf 470}, 488 (2001).
\bibitem{VPD} W. J. Llope {\it et al.} (STAR Collaboration), Nucl. Instrum. Methods A {\bf 522}, 252 (2004).
\bibitem{tof} B. Bonner {\it et al.} (STAR Collaboration), Nucl. Instrum. Methods A {\bf 508}, 181 (2003);
M. Shao {\it et al.} (STAR Collaboration), Nucl. Instrum. Methods A {\bf 558}, 419 (2006).
\bibitem{startpc} K.~H.~Ackermann {\it et al.} (STAR Collaboration), Nucl. Instrum. Methods A {\bf 499}, 624 (2003).
\bibitem{starftpc} K.~H.~Ackermann {\it et al.} (STAR Collaboration), Nucl. Instrum. Methods A {\bf 499}, 713 (2003).
\bibitem{centralitydef}
B.~I.~Abelev {\it et al.}  (STAR Collaboration), Phys. Rev. C {\bf 81}, 024911 (2010).
\bibitem{Kharzeev:2000ph} D.~Kharzeev and M.~Nardi, Phys. Lett. B {\bf 507}, 121 (2001).
\bibitem{Nakamura:2010zzi}K.~Nakamura {\it et al.}  (Particle Data Group), J. Phys. G {\bf 37}, 075021 (2010).
\bibitem{Back:2004dy}B.~B.~Back {\it et al.}  [PHOBOS Collaboration], Phys. Rev. C {\bf 70}, 021902(R) (2004).


\bibitem{recenter} J. Barrette {\it et al.} (E877 Collaboration), Phys. Rev. C {\bf 55}, 1420 (1997);
I. Selyuzhenkov and S. Voloshin, Phys. Rev. C {\bf 77}, 034904 (2008).

\bibitem{shiftMethod} J. Barrette {\it et al.} (E877 Collaboration), Phys. Rev. C {\bf 56}, 3254 (1997).
\bibitem{BBCEP} G. Agakishiev {\it et al.} (STAR Collaboration), Phys. Rev. C {\bf 85}, 014901 (2012).

\bibitem{Bilandzic:2010jr}A.~Bilandzic, R.~Snellings, S.~Voloshin, Phys. Rev. C {\bf 83}, 044913 (2011).

\bibitem{Ante_thesis}A.~Bilandzic, PhD. thesis, Nikhef and Utrecht University, 2011.
\bibitem{epart21}  J. Y. Ollitrault, A. M. Poskanzer and S. A. Voloshin, Phys. Rev. C {\bf80}, 014904 (2009).





\bibitem{ms03} M. Miller and R. Snellings, nucl-ex/0312008 (2003).
\bibitem{glauber} M. L. Miller, K. Reygers, S. J. Sanders, P. Steinberg, Ann. Rev. Nucl. Part. Sci. {\bf 57}, 205 (2007).
\bibitem{CGCa}  A. Adil {\it et al.}, Phys. Rev. C {\bf74}, 044905 (2006).
\bibitem{CGCc}  H. J. Drescher and Y. Nara, Phys. Rev. C {\bf75}, 034905 (2007).
\bibitem{CGCb}  H. J. Drescher and Y. Nara, Phys. Rev. C {\bf76}, 041903 (2007).
\bibitem{CGCd}  T. Hirano and Y. Nara, Phys. Rev. C {\bf79}, 064904 (2009).


\bibitem{eparttwo} S. Voloshin, nucl-th/0606022 (2006).
\bibitem{epart2}  B. Alver {\it et al.} (PHOBOS Collaboration), Phys. Rev. C {\bf77}, 014906 (2008).
\bibitem{star_auau62.4} B. I. Abelev {\it et al.} (STAR Collaboration), Phys. Rev. C {\bf 75}, 054906 (2007).
\bibitem{v2_4_200_62} Y. Bai, Ph.D. thesis, Nikhef and Utrecht University, 2007.
\bibitem{SPSv2}  A. M. Poskanzer and S. A. Voloshin (NA49 Collaboration), Nucl. Phys. A {\bf 661}, 341c (1999).


\bibitem{ampt} Z. Lin {\it et al.} , Phys. Rev. C {\bf 72}, 064901 (2005).
\bibitem{urqmd} H. Petersen {\it et al.}, arXiv: 0805.0567v1 (2008).
\bibitem{hadronic_potential} J. Xu {\it et al.}, Phys. Rev. C {\bf 85}, 041901 (2012).
\end{thebibliography}
\end{document}